\documentclass[aps, pra, reprint, superscriptaddress, dvipsnames]{revtex4-2}

\usepackage[dvipsnames, table]{xcolor}
\definecolor{quantumviolet}{HTML}{53257F}
\definecolor{quantumgray}{HTML}{555555}

\usepackage{multirow}

\usepackage{bm}
\usepackage{bbm}
\usepackage[retainorgcmds]{IEEEtrantools}
\usepackage{graphicx}
\usepackage{mathrsfs}
\usepackage{amsmath}
\usepackage{amsfonts}
\usepackage{amssymb}
\usepackage{times,txfonts}
\usepackage{nicefrac}
\usepackage{ragged2e} 
\usepackage{physics} 
\usepackage{verbatim} 
\usepackage{blindtext}
\usepackage{lipsum}  
\usepackage[inkscapearea=page]{svg} 
\usepackage[export]{adjustbox}
\usepackage[colorlinks=true,linkcolor=blue,urlcolor=blue,citecolor=blue,pdfusetitle]{hyperref}

\usepackage{tikz}                
\usetikzlibrary{calc}            
\usetikzlibrary{positioning}     

\usepackage{tcolorbox}

\usepackage{algorithm}
\usepackage[noend]{algpseudocode}

\usepackage{multirow} 
\usepackage{makecell} 

\definecolor{cadmiumgreen}{HTML}{097969}

\newcommand{\soliton}{\includegraphics[valign=c]{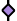}}
\newcommand{\Xdraw}{\includegraphics[valign=c]{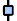}}
\newcommand{\Ydraw}{\includegraphics[valign=c]{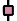}}
\newcommand{\leg}{\includegraphics[valign=c]{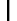}}
\newcommand{\cphasedraw}{\includegraphics[valign=c]{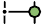}}
\newcommand{\swapdraw}{\includegraphics[valign=c]{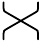}}

\newcommand{\expec}[1]{\langle #1 \rangle}
\newcommand{\swap}{\mathrm{SWAP}}

\newcommand{\cptp}{\mathcal{E}}
\newcommand{\PTr}[2]{\mathrm{Tr}_{#1} \left\{ #2 \right\}}
\newcommand{\vectorize}[1]{\mathrm{vec} \left[ #1 \right]}
\newcommand{\id}{\mathbbm{1}}
\newcommand{\phasegate}{\phi}
\newcommand{\CPhase}{\mathrm{CP}(\phasegate)}

\newcommand{\yvalidation}{{\bf y}_{\mathrm{val}}}
\newcommand{\ytarget}{{\bf y}_{\mathrm{tar}}}

\begin{document}

\title{Dual-unitary Circuits as a Platform for Quantum Reservoir Computing}
\date{\today}

\author{Gabriel O. Alves}
\email{alves.go.co@gmail.com}
\affiliation{Max Planck Institute for the Physics of Complex Systems, 01187 Dresden, Germany}

\author{Pieter W. Claeys}
\affiliation{Max Planck Institute for the Physics of Complex Systems, 01187 Dresden, Germany}
\affiliation{School of Physics, Trinity College Dublin, Dublin 2, Ireland}

\begin{abstract}

Quantum reservoir computing (QRC) is a machine learning approach which employs the internal dynamics of a physical system (the reservoir) to encode and process information. 
In this work, we explore the use of dual-unitary circuits in a brickwork architecture as a platform for QRC, well suited to current noisy intermediate-scale quantum devices. 
Dual-unitary circuits present both practical and conceptual advantages.
Our results indicate that, under appropriate conditions, dual-unitarity can lead to an enhanced regime of operation: we numerically verify that it improves memory effects and nonlinear processing, and shields against finite-shot noise, mitigating exponential concentration. 
Moreover, dual unitarity offers an intuitive picture of how operator dynamics gives rise to memory and nonlinear processing in circuit-based reservoirs.

\end{abstract}

\maketitle{}


\section{Introduction}

Reservoir computing (RC) is a supervised machine learning approach which traces back to the seminal works by Jaeger~\cite{jaegerEchoStateApproach2001} and Maass \emph{et al.}
~\cite{maassRealtimeComputingStable2002} on echo state networks and liquid state machines, respectively.
This approach is based on the use of a fixed physical system as a substrate encoding input data.
Features of this physical substrate are then read out and used to train a single linear output layer.
In comparison to more sophisticated architectures, such as deep or recurrent neural networks, RC provides the clear advantage of being much easier to train.
Due to its inherent properties, RC has successfully found applications in certain classes of tasks, such as time-series processing and the forecasting of (nonlinear) dynamical systems~\cite{nakajimaReservoirComputingTheory2021, vrugtIntroductionReservoirComputing2024, salatinoForecastingLowDimensionalTurbulence2025}.
With the advance of quantum technologies in the past three decades and the emergence of noisy intermediate-scale quantum (NISQ) devices, there has been a simultaneous surge in the search for suitable algorithms and applications for such hardware.
Quantum reservoir computing (QRC) presents a hybrid approach merging these two worlds.

\begin{figure}
    \centering
    \includegraphics[width=\columnwidth]{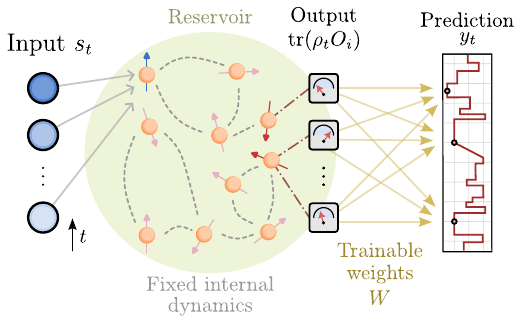}
    \caption{
    An illustration of a quantum reservoir computer.
    The data (blue), e.g. a time series $s_t$, is encoded into a system which displays intricate internal dynamics.
    By evolving, the system processes the data in a nonlinear way and with memory.
    By performing measurements on the system, features are extracted which are then used to train the weights ${\bf W}$ by fitting the target function (red) $y_t$.
    Unlike standard neural network architectures, the ``internal weights'' in the reservoir are fixed and set by the choice of reservoir, and the training procedure reduces to a linear optimization on the single output layer.
    This picture represents a simplified QRC protocol where all the input data is encoded in a single component of the system, e.g., a single qubit in a lattice.
    }
    \label{fig:QRC_diagram}
\end{figure}

Since its introduction by Fujii and Nakajima~\cite{fujiiHarnessingDisorderedEnsembleQuantum2017a, fujiiQuantumReservoirComputing2021}, QRC has gained considerable traction~\cite{ghoshQuantumReservoirProcessing2019, mujalOpportunitiesQuantumReservoir2021, tariqQuantumReservoirComputing2026}.
In crude terms, QRC is an RC protocol which uses a quantum system as the substrate, as illustrated in Fig.~\ref{fig:QRC_diagram}.
It displays many analogous properties and advantages as classical RC, and excels in similar tasks~\cite{kobayashiFeedbackDrivenQuantumReservoir2024, abbasReservoirComputingUsing2024}.
Furthermore, a promising feature of this framework is the fact that virtually all available proposals can be realized on many experimentally feasible platforms, such as atoms in an optical cavity~\cite{zhuPracticalFewatomQuantum2025, dasQuantumReservoirComputing2025}, spin chains~\cite{xiaReservoirLearningPower2022b}, photonic systems~\cite{garcia-beniScalablePhotonicPlatform2023}, digital quantum circuits~\cite{hamhoumMultivariateTimeSeries2025} and bosonic models in diverse scenarios~\cite{dudasQuantumReservoirComputing2023, llodraBenchmarkingRoleParticle2023, dudasTrainingParametricInteractions2025, llodraQuantumReservoirComputing2025}, making it a sensible candidate for a quantum technology with near-term applications.

With the recent advances in QRC, there has been an increase in our understanding of the role played by different quantum features and resources in model performance~\cite{sanniaDissipationResourceQuantum2024a, garcia-beniSqueezingResourceTime2024a, karimiRoleEntanglementQuantum2025}.
Notable examples highlight how reservoirs tuned to the onset of quantum chaos can lead to an enhanced performance~\cite{martinez-penaDynamicalPhaseTransitions2021,xiaReservoirLearningPower2022b,llodraQuantumReservoirComputing2025}.

Within this context, we propose the use of dual-unitary (DU) circuits as a platform for QRC.
Such circuits enjoy the property of being unitary in both time \emph{and} space (see Ref.~\cite{bertiniExactlySolvableManybody2025} for a review).
Dual-unitary circuits present minimal models of many-body quantum dynamics in which dynamical features can be analytically characterized. Furthermore, these features are generally more pronounced than in non-DU models: If DU circuits are chaotic, they are maximally chaotic, and if DU circuits are integrable, they are superintegrable. For this reason DU circuits present a natural playground in which the effects of dynamical features on the performance of QRC can be understood.

\begin{figure}
    \centering
    \includegraphics{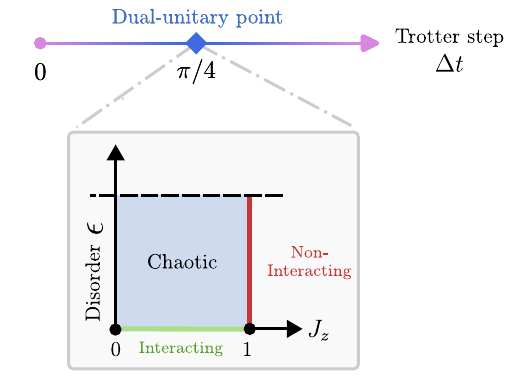}
    \caption{
    Dynamical phase diagram for the DU XXZ circuit following the parametrization in Eq.~\eqref{eq:gate_parametrization}. 
    By varying the Trotter step we can move towards and away from the DU line at $\Delta t = \pi/4$.
    The inset details the behavior of the circuit at this point.
    First, by turning off the on-site disorder with $\epsilon = 0$ (green edge), we have an interacting integrable circuit.
    The circuit displays solitons in this regime, which lead to unique behavior and memory effects, as we discuss in Sec.~\ref{sec:role_soliton_feature}.
    Meanwhile, at $J_z = 1$ the circuit is non-interacting (red edge).
    The circuit is in the chaotic regime (blue region) for $J_z \neq 1$ and nonvanishing on-site disorder ($\epsilon>0$).
    }
    \label{fig:phase_diagram}
\end{figure}

Our work takes a step in this direction: we investigate the role played by dual-unitarity in the operation of the reservoir, and how it affects memory and nonlinear capabilities.
Our numerics indicate that dual-unitarity can lead to observable effects in the operation of the reservoir.
These models display a rich collection of different dynamical regimes, as we illustrate in Fig.~\ref{fig:phase_diagram}.
A particular point of focus in this work is the interacting integrable regime (green edge in Fig.~\ref{fig:phase_diagram}).
In this regime, DU circuits are known to feature solitons~\cite{bertiniOperatorEntanglementLocal2020, holden-dyeFundamentalChargesDualunitary2025}. Note that in this context the term is used differently from its conventional meaning in the condensed matter literature: here, solitons denote operators which, upon unitary conjugation, simply move around the circuit.
This property can be used to explicitly show how solitons lead to memory effects.
The information not encoded in solitons is (maximally) scrambled, a signature of quantum chaos and responsible for the nonlinear capabilities of QRC. 
These circuits hence also serve as a pedagogical example to illustrate how operator growth is responsible for the capture of temporal correlations and nonlinearities.

We first lay the groundwork of this manuscript by introducing dual-unitary XXZ circuits (Sec.~\ref{sec:DU}), including the graphical notation we will use throughout this manuscript.
Afterwards, we briefly review the quantum reservoir computing framework based on the erase-input protocol (Sec.~\ref{sec:QRC}).
For convenience, we show how to express the erase-input map in graphical notation (Sec.~\ref{sec:QRC_DU}) and go over the basic tasks which serve as useful benchmarks for QRC (Sec.~\ref{sec:QRC_tasks}).
Then, in Sec.~\ref{sec:QRC_DU_results} we discuss our results, where we investigate how different aspects of the reservoir impact its performance.
Namely, we probe 
(i) the role of dual-unitarity and the Trotter step (Secs.~\ref{sec:role_DU} and~\ref{sec:optimality_DU}) and
(ii) the role of entanglement (Sec.~\ref{sec:benchmark_entanglement}).
We close this part of the investigation by exploring memory and nonlinearity: 
(iii) we explain the mechanism behind how solitons behave under the erase-input map, working as memory mechanisms (Sec.~\ref{sec:role_soliton_feature}), and (iv) further explore the nonlinear processing capacity of such circuits through Legendre polynomials in Sec.~\ref{sec:nonlinearity_legendre}. 
We observe a trade-off between linear memory capacity and nonlinearity, akin to classical reservoirs~\cite{dambreInformationProcessingCapacity2012}.
Our analysis also serves as a pedagogical illustration of how exactly operator dynamics leads to the recovery of nonlinear effects and temporal correlations of the input (Sec.~\ref{app:operator_growth}). 
In Sec.~\ref{sec:exp_concentration} we explore the phenomenon of exponential concentration under the lens of DU circuits.
We show that dual-unitarity and integrability might offer a certain degree of protection against exponential concentration.
Although their behavior in terms of scaling is qualitatively the same as generic circuits, dual-unitarity provides an improvement associated with a significant  prefactor over finite-shot fluctuations. 


\section{Dual-unitary XXZ circuits}
\label{sec:DU}

In this section we introduce the graphical notation and the main parametrization we will use throughout the rest of the manuscript.

\subsection{General DU circuits}

In this paper we focus on circuits constructed from unitary gates $V$ obtained from the Trotterization of the XXZ spin chain:
\begin{equation}\label{eq:XXZ_parametrization}
    V
    = 
    \exp\left[-i \Delta t
    \left( X \otimes X 
    +  Y \otimes Y 
    + J_z\, Z \otimes Z 
    \right)\right],
\end{equation}
with $X,Y,Z$ being the Pauli matrices. 
Additionally, we have two hyperparameters: $J_z$ as the anisotropy parameter, and $\Delta t \in \mathbb{R}$ as the Trotter step.
This unitary leads to an integrable circuit for any choice of parameters~\cite{vanicatIntegrableTrotterizationLocal2018, claeysErgodicNonergodicDualUnitary2021a}.
We will study the effect of breaking integrability by introducing on-site U($2$) gates $u_\pm$ through a third hyperparameter $\epsilon$ by:
\begin{equation}\label{eq:gate_parametrization}
    U(\epsilon, J_z, \Delta t) 
    =
    (u_+ \otimes u_-)
    V.
\end{equation}
The on-site unitary gates are defined as $u_\pm = e^{-i \epsilon H_\pm}$, where $H_\pm$ are $2 \times 2$ GUE random matrices~\cite{livanIntroductionRandomMatrices2018a}.
The parameter $\epsilon$ tunes the on-site disorder, leading to a nonintegrable (chaotic) model for any $\epsilon \neq 0$, and setting $\epsilon = 0$ turns off the single-site gates.
Unless expressed otherwise, we will keep disorder homogeneous throughout the circuit.
More importantly, at $\Delta t = \pi/4$ the gate is dual-unitary for any choice of $J_z$ and $\epsilon$.
We emphasize that the model can be dual-unitary irrespective of whether it is integrable or chaotic.

The gate in Eq.~\eqref{eq:gate_parametrization} constitutes the building block of our circuit, and hence the reservoir, being graphically represented as:
\begin{equation}\label{eq:main_circuit_graphic}
    U = \includegraphics[valign=c]{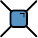}\,.
\end{equation}
We consider a one-dimensional lattice of $L$ qubits with open boundary conditions. The so-called Floquet unitary $U_F$ describing a single discrete time step is given by a product of an even and odd layer as follows:
\begin{equation}\label{eq:floquet_circuit}
    U_F = U_e U_o = (\id \otimes U \otimes ... \otimes U)(U \otimes ... \otimes U \otimes \id).
\end{equation}
Graphically, this translates to:
\begin{equation}\label{eq:floquet_circuit_drawn}
    U_F = \includegraphics[valign=c]{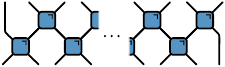}.
\end{equation}
Without any loss of generality, we assume odd circuit size $L$ throughout the manuscript, as depicted in the equation above. 
In Fig.~\ref{fig:phase_diagram} we provide a dynamical phase diagram for the Floquet circuit above, given the parametrization Eq.~\eqref{eq:gate_parametrization}.
We will structure our analysis in this manuscript around this diagram.
Namely, we will investigate how the performance of the reservoir and its underlying features change across its different regions.

\subsection{Solitons and integrable DU circuits}
\label{sec:soliton_def}

We first discuss some important properties of the interacting integrable circuit ($\epsilon = 0$, no on-site disorder) at the dual-unitary point $\Delta t = \pi/4$.
In this case, we can re-express the gate from Eq.~\eqref{eq:XXZ_parametrization} as $V = \CPhase \swap$, where
\begin{equation}
    \swap  = \swapdraw    
\end{equation}
is the swap gate and
\begin{equation}\label{eq:cphase_gate}
    \CPhase 
    = 
    \ketbra{0}{0} \otimes \id +  \ketbra{1}{1} \otimes \left(\ketbra{0}{0} + e^{i \phi}\ketbra{1}{1}\right) 
    =:
    \cphasedraw
\end{equation}
is a controlled-phase gate. 
The relation to the dual-unitary parametrization in Eq.~\eqref{eq:XXZ_parametrization}, evaluated at $\Delta t=\pi/4$, is obtained by writing
\begin{equation}
V(J_z)=e^{-i\pi J_z/4}(P\otimes P)\CPhase\swap,
\end{equation}
with $P=\operatorname{diag}(1,e^{-i\phi/2})$ and $\phi=\pi(1-J_z)$~\cite{suzukiComputationalPowerOne2022b}.
In this limit the Floquet circuit can be depicted as
\begin{equation}\label{eq:circuit_int}
    U_F
    =
    \includegraphics[valign=c]{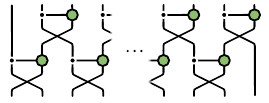}.
\end{equation}

The presence of solitons in this regime will allow us to make some statements about the memory effects in the reservoir.
As mentioned previously, in the jargon of DU circuits solitons are traceless operators which are translated along the lattice under the circuit dynamics~\cite{bertiniOperatorEntanglementLocal2020}.
For the gate $V$ appearing in the circuit of Eq.~\eqref{eq:circuit_int}, the solitons correspond to the Pauli $Z$-matrices.
Given the graphical representation,
\begin{equation}\label{eq:soliton_graphical_definition}
    Z_x
    =:
    \hdots
    \underset{x - 1}{\leg}
    \:
    \underset{x}{\soliton}
    \:
    \underset{x + 1}{\leg}
    \hdots,
\end{equation}
with $Z_x$ being the Pauli matrix $Z$ at site $x$, and $Z$ acting on the computational basis states as $Z\ket{0}=-\ket{0}$ and $Z\ket{1}=\ket{1}$, we have
\begin{equation}
    V^\dagger (\soliton \leg) V
    =
    V^\dagger (Z \otimes \id) V
    =
    (\id \otimes Z) 
    =
    \leg \soliton. 
\end{equation}
Analogously, we have $ V^\dagger (\leg \soliton) V =  (\soliton \leg)$.

\subsection{Folded representation}

Given the density matrix evolution $\rho_t = U_F \rho_{t-1} U_F^\dagger$, we can write this in the so-called folded (or replica) picture by means of vectorization (see Lemma 1 of Ref.~\cite{gilchristVectorizationQuantumOperations2011}, for example).
More specifically, we have $\vectorize{\rho_t} = (U_F^* \otimes U_F) \vectorize{\rho_{t-1}}$.
The unitary evolution is thus represented through the \emph{folded gates}:
\begin{equation}
    U^* \otimes U 
    = 
    \includegraphics[valign=c]{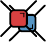}
    =:
    \includegraphics[valign=c]{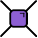}    
    .    
\end{equation}
Meanwhile, the unitarity condition reads:
\begin{equation}\label{eq:folded_unitarity}
    \includegraphics[valign=c]{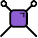}
    =
    \includegraphics[valign=c]{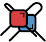}
    =
    \includegraphics[valign=c]{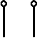}    
    .    
\end{equation}
where circles denote the vectorization of $\id$.
Additionally, it proves to be convenient to graphically denote the interacting integrable gate, also in the folded picture, by:
\begin{equation}
    V^* \otimes V
    =:
    \includegraphics[valign=c]{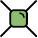}    
    .    
\end{equation}
For visual clarity, we distinguish the integrable folded gate by using a curved wedge, in contrast to the straight wedge of the generic gate.
In this notation we can express the translation of solitons as follows:
\begin{equation}\label{eq:soliton_folded}
    \vectorize{V^\dagger(Z \otimes \id)V}
    =
    \includegraphics[valign=c]{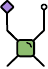}  
    =
    \includegraphics[valign=c]{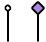}  
    =
    \vectorize{\id \otimes Z},
\end{equation}
with an analogous representation for the left-moving solitons.


\section{Quantum Reservoir Computing}
\label{sec:QRC}

In this section we give a brief overview of the QRC paradigm.
In particular, we implement the erase-input protocol proposed by Fujii and Nakajima in their seminal work on QRC~\cite{fujiiHarnessingDisorderedEnsembleQuantum2017a}.
Additionally, we will lay everything out in graphical notation for clarity, since this language will be important later on in Sec.~\ref{sec:role_soliton_feature} when discussing solitons.
As we illustrate in Fig.~\ref{fig:QRC_diagram}, the reservoir computing protocol can be divided into roughly three phases:

\begin{tcolorbox}[colback=RoyalBlue!10!white,colframe=RoyalBlue!75, title={\centering \textbf{QRC protocol}}]

\begin{enumerate}
    \item \textbf{Encoding phase.} 
        The input data is physically encoded into the reservoir, influencing and driving its dynamics. 
        In QRC this is usually done at the level of quantum states, regardless of the specific choice of platform.
        In the erase-input protocol, the encoding is done as defined in Eq.~\eqref{eq:erase_input}.
    \item \textbf{Dynamics and readout phase.}
        Between successive inputs, the reservoir evolves under its own internal dynamics, which spreads the encoded information across its degrees of freedom.
        We then perform a readout of the chosen observables, whose expectation values constitute the \emph{features} of the model.
        This results in a feature matrix whose dimensionality depends on the number of time steps and the number of observables [Eq.~\eqref{eq:features_graphical}].
    \item \textbf{Training and prediction. }
        Once the features are obtained, we perform a linear optimization to find the weight vector that best maps the features to the target outputs.
        A typical choice is a simple linear regression algorithm, such as Ridge regression [Eq.~\eqref{eq:ridge}].
        The trained weights can then be used to perform predictions.
\end{enumerate}

\end{tcolorbox}

Suitable candidates for QRC generally feature some desirable properties, such as the echo state property, which is related to the forgetting of initial conditions~\cite{yildizRevisitingEchoState2012}, and the ability to provide nonlinearity~\cite{goviaNonlinearInputTransformations2022,mujalAnalyticalEvidenceNonlinearity2021}.
It was also shown that dissipation is a useful and, under certain conditions, \emph{necessary} resource for QRC~\cite{sanniaDissipationResourceQuantum2024a}.
The erase-input map of Fujii and Nakajima corresponds to a CPTP map satisfying these criteria~\cite{martinezpenaInputdependenceQuantumReservoir2025a}.
For a useful and pedagogical introduction to classical RC instead, see Refs.~\cite{cucchiHandsonReservoirComputing2022, wringeReservoirComputingBenchmarks2025}.

\subsection{QRC with quantum circuits}
\label{sec:QRC_DU}

We now discuss the erase-input protocol in detail.
Our focus in this section will be to translate the relevant equations into tensor-network language for convenience.
We will once again proceed with vectorized notation and write, for instance, $\Tr{AB} = \vectorize{A}^\dagger\vectorize{B}$ ~\cite{watrousTheoryQuantumInformation2018}.
In this picture, the erase-input phase of the protocol, which corresponds to the encoding phase, is described by a completely positive trace-preserving (CPTP) map of the form:
\begin{equation}\label{eq:erase_input}
    \cptp(\rho_{t - 1})
    := 
    \ketbra{s_t}{s_t} \otimes \PTr{1}{\rho_{t-1}}.
\end{equation}
Here, we perform the temporal encoding of the data by tracing out, i.e. ``erasing'' the first qubit, and then ``inserting'' a new one at the same place through the superposition state:
\begin{equation}
    \ket{s_t} = \sqrt{1 - s_t}\ket{0} +  \sqrt{s_t}\ket{1}, \quad s_t \in [0, 1].
\end{equation}
We represent the injected state as:
\begin{equation}
    \vectorize{\ketbra{s_t}{s_t}}
    =:
    \includegraphics[valign=c]{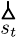}.
\end{equation}
%
The (folded) normalization condition reads
\begin{equation}
|\braket{s_t}{s_t}|^2 =
\includegraphics[valign=c]{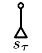} = 1.
\end{equation}
Similarly, we represent density matrices as:
\begin{equation}
    \vectorize{\rho_t}
    =:
    \includegraphics[valign=c]{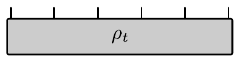}
\end{equation}
Graphically, we can rewrite the CPTP map above as:
\begin{equation}\label{eq:erase_input_graphical}
    \vectorize{\cptp(\rho_{t - 1})}
    =: \quad
    \includegraphics[valign=c]{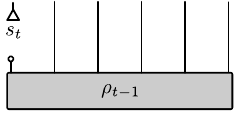}
\end{equation}
We then evolve the system unitarily, leading us to the composite stroboscopic map $\rho_t=:\Phi(\rho_{t-1})$, with
\begin{equation}
    \Phi(\rho_{t-1})=U_F\, \cptp(\rho_{t - 1}) U_F^\dagger
\end{equation}
which describes the dynamics under the erase-input protocol.
The unitary in this map is given by the Floquet circuit we introduced in Eq.~\eqref{eq:floquet_circuit_drawn}.
Graphically, it can be represented by
\begin{equation}\label{eq:cptp_map_graphic}
    \vectorize{\Phi(\rho_{t-1})}
    = \quad
    \includegraphics[valign=c]{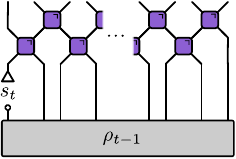}
\end{equation}
for the brickwork Floquet circuit.

Finally, we extract the output $x_{t,j}$ by computing the expected value of the observables $\{O_j\}$ at time $t$ with $x_{t, j} = \mathrm{tr}\{\rho_t O_j\} = \vectorize{O_j}^\dagger\vectorize{\rho_t}$, where $j = 1, ..., M$ indexes the $M$ features, that is, the number of observables we measure at each time step in the so-called \emph{readout} phase.
The outcomes $x_{t, j}$ constitute the entries of the so-called \emph{feature matrix}.
Graphically, we can represent these entries as:
\begin{equation}\label{eq:features_graphical}
    x_{t, j} 
    = 
    \mathrm{tr}\{\rho_t O_j\}
    =
    \includegraphics[valign=c]{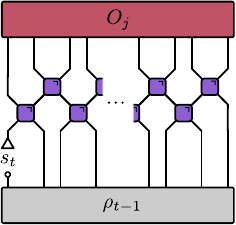}
\end{equation}
We stress that although this protocol is widespread in QRC, there are other viable alternatives~\cite{sanniaDissipationResourceQuantum2024a}.
The encoding part, for instance, can be directly implemented as parameters in the model Hamiltonian~\cite{settinoMemoryAugmentedHybridQuantum2024, mccaulMinimalQuantumReservoirs2025}, as opposed to being performed on the qubit.

\begin{figure}
    \centering
    \includegraphics[width=\columnwidth]{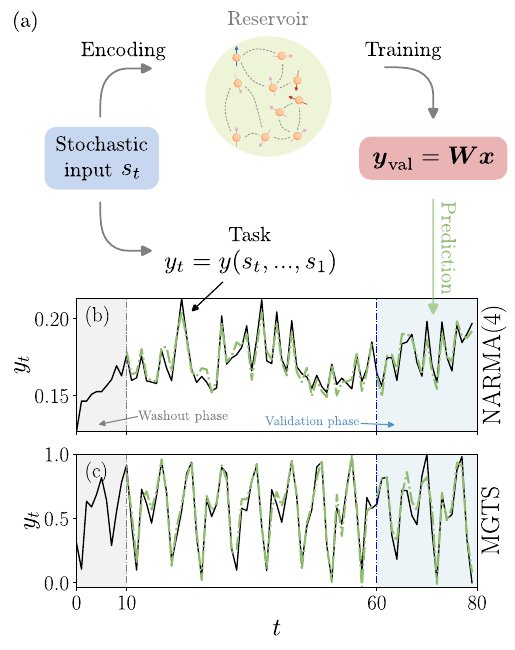}
    \caption{
    (a)
    The stochastic input $s_t$ is fed into Eq.~\eqref{eq:erase_input_graphical} to encode its value into the system; simultaneously, it also serves as an argument for the chosen benchmark $y_t$ from Sec.~\ref{sec:QRC_tasks} [any of Eqs.~\eqref{eq:stm_task}-\eqref{eq:glass_task}].
    The feature matrix $\boldsymbol{x}$ is computed through Eq.~\eqref{eq:features_graphical}, which allows for the calculation of the weight vector $\textbf{W}$ in Eq.~\eqref{eq:ridge}.
    Finally, the validation results are calculated through Eq.~\eqref{eq:prediction}, where we try to fit the target time-series $y_t$.
    We plot some basic numerical results to illustrate the framework,
    where we compare the analytical time-series (solid black line) with the numerical predictions (dashed green line), for both the (b) NARMA($4$) and 
    (c)
    the MGTS tasks.
    The data is split into three sets.
    First the washout phase (gray), then the training phase (white), and finally the validation phase (blue).
    }
    \label{fig:time_series_example_diagram}
\end{figure}

Before the expectation values are actually included in the feature matrix, one normally lets the system evolve for $t_\mathrm{washout}$ steps in what we call the \emph{washout phase}. 
In this phase, the readout step is skipped altogether and no measurements are performed (see the gray region in Fig.~\ref{fig:time_series_example_diagram}).
The purpose of this step is to eliminate any dependence on the initial state.

Once the output data is obtained, we can perform a linear optimization on $y_t = \sum_j w_{j} x_{t, j}$ in order to obtain the \emph{weight} vector $\bf W$.
One usually proceeds by adopting the Ridge regression with parameter $\alpha$, given by:
\begin{equation}\label{eq:ridge}
    {\bf W}
    =
    ({\bf x}^T{\bf x} + \alpha {\bf I})^{-1}{\bf x}^T
    {\bf y}.
\end{equation}
The bias term is fitted separately by the solver and is not affected by the regularization.
This is the final step of the protocol, known as the \emph{training phase}.
Hereafter, we take $\alpha = 5\times10^{-3}$.
If we consider the dynamics over $T$ time steps, ${\bf x} = {\bf x}({\bf s})$ is the feature matrix of dimensions $T \times M$, which is fully determined by the inputs ${\bf s}$.
Given the vector of validation inputs ${\bf s}_{\mathrm{val}}$ and its associated feature matrix ${\bf x}_{\mathrm{val}}$, the prediction is given by:
\begin{equation}\label{eq:prediction}
    \yvalidation
    =
    {\bf x}_{\mathrm{val}} {\bf W} + {\bf b},
\end{equation}
where ${\bf b}$ is the explicit bias term fitted by the solver~\cite{sklearn-ridge-docs}.
Once the training is complete, the performance is quantified through the information processing capacity (IPC) $C$, which is given by the square of the Pearson correlation coefficient between the predicted output $\yvalidation$ and the target function $\ytarget$ corresponding to the ``true'' value:
\begin{equation}\label{eq:capacity_definition}
    C
    =
    \frac{\mathrm{cov}^2(\yvalidation, \ytarget)}
    {\mathrm{var}(\yvalidation)\mathrm{var}(\ytarget)},
\end{equation}
which lies between zero and one.
We summarize the protocol in Fig.~\ref{fig:time_series_example_diagram}~(a).

\subsection{QRC tasks}
\label{sec:QRC_tasks}

We now introduce a few common tasks in RC which will serve for benchmarking purposes.

\subsubsection{Short-term memory}

The short-term memory (STM) task is the simplest target function which we will consider. 
It corresponds to the time-delayed series
\begin{equation}\label{eq:stm_task}
    y_t = s_{t - \tau}^d,
\end{equation}
where $s_t$ is sampled from the uniform distribution between $0$ and $1$.
The objective in the STM task is to recover the input from $\tau$ steps in the past, $s_{t - \tau}$.
For $d = 1$ this is a benchmark of linear memory for the reservoir and by taking $d > 1$ we introduce some degree of nonlinearity to it.
Note, however, that there are some caveats in regards to this choice. 
We reserve this discussion to Sec.~\ref{sec:nonlinearity_legendre}.

\subsubsection{Parity Check}

The parity check task is given by the target function
\begin{equation}\label{eq:parity_task}
    y_t = \sum_{k = 0}^\tau s_{t - k} \mod 2,
\end{equation}
where $s_t \in \{0, 1\}$ is a binary random variable.
This is a highly nonlinear function, which is used to evaluate the capacity of the reservoir to treat nonlinear data.
Analogously to the previous task, we can again tune $\tau$.

\subsubsection{NARMA}

The normalized auto-regressive moving average,  NARMA($\tau$), task is a time-series introduced in Ref.~\cite{atiyaNewResultsRecurrent2000}, where $\tau$ denotes its degree.
For a good fit, the reservoir needs to be able to capture both nonlinearity and short-term memory.
With the usual choice of parameters we have:
%
%
\begin{equation}\label{eq:narma_task}
    y_t = 0.3\, y_{t-1} + 0.05\, y_{t-1} \left( \sum_{i=1}^\tau y_{t-i} \right) + 1.5\, s_{t-\tau} s_{t-1} + 0.1.
\end{equation}
For this task, we take the interval $s_t \in [0, 0.2)$
\footnote{If this interval is too large, the NARMA series might diverge.}.

\subsubsection{Sine-square task}

\begin{figure}
    \centering
    \includegraphics[width=\columnwidth]{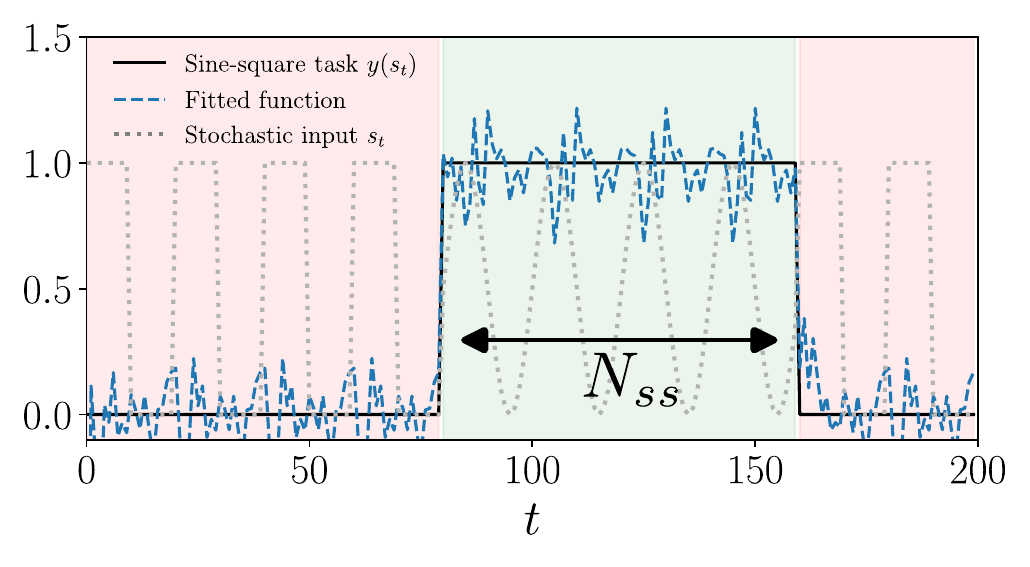}
    \caption{
    An illustration of the sine-square task. The gray dotted line depicts the stochastic inputs $s_t$, which belong either to a square or sine wave. 
    The corresponding classification $y_t = y(s_t) \in \{0, 1\}$, given by Eq.~\eqref{eq:sine-square}, is drawn with the solid black line.
    As a further guide to the eye, we also shade these regions with red and green colors on the background.
    The value $N_{ss} = 20$ is the number of points per packet.
    Finally, the blue dashed line displays an example for the fitted function.
    }
    \label{fig:sine_square_diagram}
\end{figure}

Additionally, we also introduce a basic classification task for a sine-square waveform \cite{paquotOptoelectronicReservoirComputing2012, dudasQuantumReservoirComputing2023, zhuMinimalisticScalableQuantum2025}.
In this task, sine or square wave-packets are randomly concatenated with each other, with each packet consisting of $N_{ss}$ discrete points. 
The task corresponds to determining whether a given point $s_t \in [0, 1]$ belongs either to a sine or square packet.
The target function can thus be written as:
\begin{equation}\label{eq:sine-square}
    y_t
    =
    \begin{cases}
    0 & \text{if } s_t \in \text{square wave} \\
    1 & \text{otherwise}
    \end{cases}
\end{equation}
Given a sequence of points $s_t$ in a wave, our task is to classify it according to the equation above, as illustrated in Fig.~\ref{fig:sine_square_diagram}.

\subsubsection{Mackey-Glass time series}

For completeness we also include the Mackey-Glass time series, defined through the differential equation:
\begin{equation}\label{eq:glass_task}
    \frac{dx(t)}{dt}
    =
    \beta \frac{x(t-\kappa)}{1 + x(t-\kappa)^n} - \gamma x(t).
\end{equation}
with the typical parameter values $\beta = 0.2$, $\gamma = 0.1$, $n = 10$, and $\kappa = 17$, where the model is chaotic~\cite{mackeyOscillationChaosPhysiological1977,
fryOptimizingQuantumNoiseinduced2023a}.
In the numerics, we obtain a discrete input sequence by retaining every tenth point of the resulting trajectory, and each realization is normalized to obtain the input sequence $s_t \in [0,1]$.
For this task, we use the discretized time series as a forecasting task, where the reservoir is trained to predict the value $y_t = s_{t+\tau}$ at $\tau$ steps ahead.
Examples of realizations of the NARMA series and MGTS are shown in Fig.~\ref{fig:time_series_example_diagram}~(b) and (c), respectively.


\section{DU circuits as a QRC platform}
\label{sec:QRC_DU_results}

In this section we discuss our main results for dual-unitary and integrable circuits.
By exploring the parameter space of the model, according to the parametrization given in Eq.~\eqref{eq:gate_parametrization}, we investigate the role played by different aspects of the reservoir.
We tune  integrability and the degree of entanglement in the circuit through the (i) local disorder $\epsilon$ and (ii) anisotropy parameters $J_z$ , respectively.
Similarly, we also vary (iii) the Trotter step $\Delta t$ to tune the model to and away from dual-unitarity.
These aspects are numerically discussed in Secs.~\ref{sec:role_DU},~\ref{sec:optimality_DU} and~\ref{sec:benchmark_entanglement}.
For convenience, we provide a summary of results in Table~\ref{table:summary}.

\begin{table}[]
    \begin{tcolorbox}[colback=RoyalPurple!10!white,colframe=RoyalPurple!75, title={\centering \textbf{SUMMARY OF RESULTS}}]
    \begin{itemize}
        \item \textbf{Memory effects in the circuit.} 
            We investigate how dual-unitarity and ergodicity affect the propagation of information and the memory capabilities in different tasks [Figs.~\ref{fig:tasks_vs_delay} and ~\ref{fig:tasks_vs_trotter}].  
            Depending on the task and choice of parameters, dual-unitarity displays improved memory effects for time-lags smaller than system size.
            
            %
            
        \item \textbf{Role of entangling power.}
            Numerically, we generally find a non-monotonic behavior in which a moderate,  but not maximal, amount of entanglement is beneficial [Fig.~\ref{fig:tasks_vs_entanglement}].
        \item \textbf{Operator growth.}
            We show how operator dynamics acts as a simple memory and nonlinear-processing mechanism in the dual-unitary case. While we focus on the integrable case for concreteness, these results extend to ergodic dynamics [Figs.~\ref{fig:tasks_vs_feature_DU} and ~\ref{fig:tasks_vs_feature_DU_parity_3D}].
        \item \textbf{Effects on exponential concentration. }
            We show numerical examples of how dual-unitarity and integrability, as well as the local structure of brickwork circuit, protect the system against the effect of exponential concentration and finite-shot noise [Figs.~\ref{fig:capacity_vs_noise} and ~\ref{fig:concentration}].

    \end{itemize}
    \end{tcolorbox}
    \caption{Summary of results.}
    \label{table:summary}
\end{table}
We take the IPC [Eq.~\eqref{eq:capacity_definition}] as the main metric of performance throughout the discussion.
In some plots we colloquially refer to the ``error" as $1 - C$ in order to improve the visualization, showing some curves in log-scale.
We take our set of features to be given by the single-body Pauli operators $\{X_j, Y_j, Z_j\}_{j=1,..,L}$, the set of nearest-neighbor operators $\{X_j X_{j + 1}, Y_j Y_{j + 1}, Z_j Z_{j + 1}\}_{j=1,..,L - 1}$ and the set of next-to-nearest-neighbor operators $\{X_j X_{j + 2}, Y_j Y_{j + 2}, Z_j Z_{j + 2}\}_{j=1,..,L - 2}$, totaling $9(L - 1)$ observables (nodes per Fig.~\ref{fig:QRC_diagram}).
We let the system evolve for $t_\mathrm{washout} = 40L$ steps in the washout phase.
The training set corresponds to $t_\mathrm{training} = 80L$ points, while we consider $t_\mathrm{val} = 20L$ time steps to validate the data.
We keep these choices fixed unless expressed otherwise.

\subsection{Role of dual-unitarity and integrability}
\label{sec:role_DU}

\begin{figure}
    \centering
    \includegraphics[width=0.95\columnwidth]{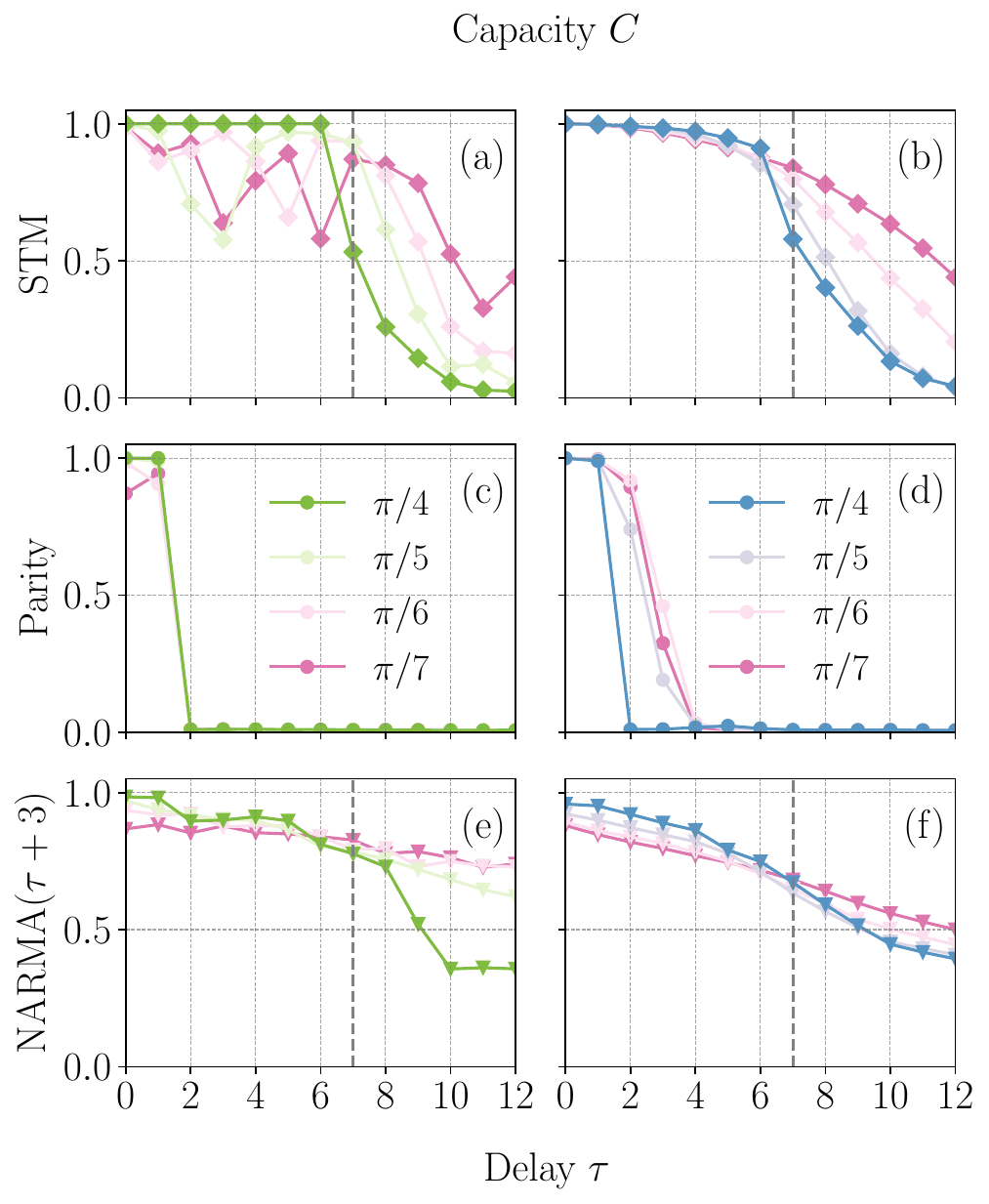}
    \caption{
        Information processing capacity as a function of the delay for: 
        [top row, panels (a) and (b)] the STM task [Eq.~\eqref{eq:stm_task}], 
        [middle row, panels (c) and (d)] the parity check task [Eq.~\eqref{eq:parity_task}], and
        [bottom row, panels (e) and (f)] the NARMA($\tau + 3$) task [Eq.~\eqref{eq:narma_task}].
        We plot the results for the integrable (left column, $\epsilon = 0$) and ergodic regimes (right column, $\epsilon=1$), for four values of the Trotter step, $\Delta t = \pi/4$ (the dual-unitary point) through $\pi/7$, indicated by a color gradient from green/blue (at $\Delta t = \pi/4$) to pink (away from dual-unitarity); marker shapes indicate the task (diamonds: STM, circles: parity check, triangles: NARMA). The dashed vertical line marks $\tau = L$.
        The presence of solitons in the integrable case leads to perfect (linear) memory up to $\tau=L - 1$ for the STM, with a sharp decline for delays equal to or larger than system size.
        The simulations were performed for $L = 7$ and $50 \times 100 = 5000$ circuit and task realizations, respectively.
    }
    \label{fig:tasks_vs_delay}
\end{figure}

We start our discussion by plotting the IPC for the short-term memory, parity check and NARMA tasks as a function of the delay $\tau$ in Fig.~\ref{fig:tasks_vs_delay}.
We show our results in the integrable (left column) and ergodic (right column) configurations, for different values of Trotter step $\Delta t$.
All figures indicate a similar result: For sufficiently small delay $\tau$, the dual-unitary gates outperform the non-dual-unitary gates, whereas for larger delays the dual-unitary gates are outperformed (or perform similar to) the non-dual-unitary gates.

This feature is most pronounced for the integrable circuit at the dual-unitary point, where we have perfect accuracy for the linear STM task up to delay $\tau = L - 1$, as seen in panel (a). 
As we later discuss (Sec.~\ref{sec:role_soliton_feature}), this is entirely explained by the transport of solitons in the circuit.
We show that in the integrable regime the solitons work as a perfect propagator of linear memory in the circuit.
We then see a sharp drop-off in performance for $\tau \geq L$, which happens for the same reason: due to the erase part of the map [Eq.~\eqref{eq:erase_input}], the solitons are unable to convey any information for time-lags larger than system size.
Remarkably, signatures of this feature persist to the ergodic case, see panel (b), where the dual-unitary configuration again outperforms the others for $\tau < L$ (see again Sec.~\ref{sec:role_soliton_feature}). We observe the same crossover for delays equal to or larger than system size, where generic non-DU configurations (pink line) overtake and display improved memory.

For the parity in the middle row [Figs.~\ref{fig:tasks_vs_delay}~(c) and (d)], we see a sharp drop-off both for the integrable circuit and for the DU chaotic circuit for $\tau > 1$. This task requires highly non-local features in order to properly recover the nonlinearity in Eq.~\eqref{eq:parity_task}, and these results indicate that the circuit cannot reproduce such non-local features. 
This can again be shown analytically in the integrable DU case (see Sec.~\ref{app:operator_growth}). 
Overall, the performance in such fine-tuned integrable regimes is strongly dependent on which features are available (as we later discuss in Sec.~\ref{sec:role_soliton_feature} and Fig.~\ref{fig:tasks_vs_feature_DU_parity_3D}). 
In the ergodic case, conversely, the maximal scrambling of DU circuits indicates that the targeted non-local features cannot be recovered. 
For generic unitary circuits, on the other hand, such information can be more easily retrieved due to the slower scrambling of information across the circuit~\cite{nahumOperatorSpreadingRandom2018}.

In the last row we plot the results for the NARMA($\tau + 3$) task, given by Eq.~\eqref{eq:narma_task} [Fig.~\ref{fig:tasks_vs_delay}~(e) and (f)]. 
The NARMA task corresponds to a more complicated time series, where both temporal correlations and nonlinearities come into play.
Overall, letting the gate operate around the DU configuration seems to provide reasonable performance in most scenarios, up to an eventual crossover around NARMA order $L + 3$, where the non-DU counterpart starts to perform better.
This crossover is especially evident in the chaotic case, as we show in Fig.~\ref{fig:tasks_vs_delay}~(f).
Since the nonlinearities here are less extreme when compared to the parity check task, the NARMA task does not seem to suffer as much with our choice of local features, even in the DU case.
The behavior here is qualitatively much closer to what we see in panels (a) and (b) for the STM task, but with a less stark drop-off for longer times.

\begin{figure}
    \centering
    \includegraphics[width=0.95\columnwidth]{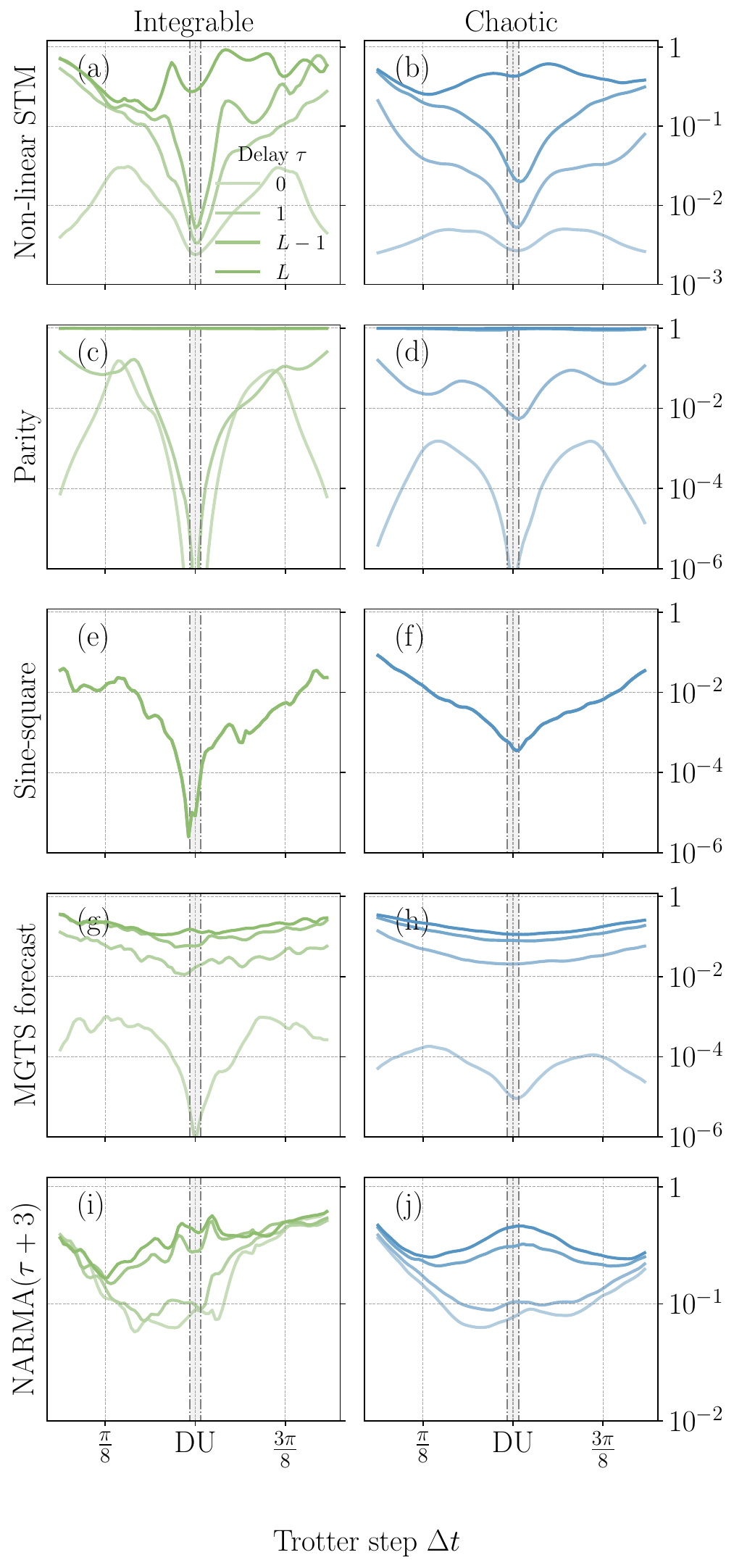}
    \caption{
        Error (given by the "complement" $1 - C$ of the IPC) as a function of the Trotter step for the nonlinear STM with $d = 2$ [(a), (b) -- Eq.~\eqref{eq:stm_task}], parity check [(c), (d) -- Eq.~\eqref{eq:parity_task}], sine-square [(e), (f) -- Eq.~\eqref{eq:sine-square}], MGTS [(g), (h) -- Eq.~\eqref{eq:glass_task}] and NARMA($\tau
        + 3$) [(i), (j) -- Eq.~\eqref{eq:narma_task}] tasks.
        We highlight the region around the dual unitary point $\Delta t = \pi/4$ with the gray area.
        Results are shown for the integrable (left column, $\epsilon = 0$) and ergodic regimes (right column, $\epsilon=1$), in analogy to Fig.~\ref{fig:tasks_vs_delay}.
        Note the prominent dip at the DU point for delays up to  $\tau = L - 1$ for most tasks, indicating the optimality around the DU point, corresponding to the Trotter step $\Delta t = \pi/4$.
        Note that in the case of the sine-square [(e), (f)] we plot a single curve, since there is no notion of delay or order in this task.
        The results were averaged over $50 \times 40 = 2000$ task and gate realizations, respectively, with $J_z = 0.8$ fixed.
        In this plot we have set $L = 5$.
        We take $40L$ points for the washout phase. 
    }
    \label{fig:tasks_vs_trotter}
\end{figure}

\subsection{Optimality near the dual-unitary point}
\label{sec:optimality_DU}

In order to probe the apparent optimality of dual-unitarity for a short delay in more detail, we show an analogous plot in Fig.~\ref{fig:tasks_vs_trotter}, now also including the sine-square and MGTS tasks alongside STM, parity check and NARMA.
We compute the error $1-C$ versus the Trotter step $\Delta t$ tuning the distance from dual-unitarity (note the log scale in the vertical axis).
In these panels, darker curves indicate a longer delay $\tau$.
We highlight the region around the DU point with $\Delta t = \pi/4$ as a guide to the eye.
We can see that for delays smaller than system size the IPC generally increases as we approach the DU point for all tasks we consider. 
As previously mentioned, in the (nonlinear) STM task [Figs.~\ref{fig:tasks_vs_trotter}~(a) and (b)], for instance, we observe this behavior up to $\tau = L - 1$.
Meanwhile, for delays larger than system size we might observe the converse behavior.
In this situation the optimal point shifts to a configuration which is \emph{not} dual-unitarity.

In the parity check tasks [Figs.~\ref{fig:tasks_vs_trotter}~(c) and (d)] we only observe good performance for $\tau = 0$ and $\tau = 1$, again with a minimum at the DU point.
This happens due to the fact that we only consider local features, a limitation discussed in more detail in Sec.~\ref{sec:role_soliton_feature}.
At the DU point the integrable configuration leads to better performance for $\tau < L$.
Away from the DU point, however, ergodicity can lead to improvements.

For the sine-square task, we plot a single curve, as there is no meaningful notion of delay.
The results seem to indicate that both dual-unitarity and ergodicity lead to better performance, as seen in Figs.~\ref{fig:tasks_vs_trotter}~(e) and (f).
We observe essentially the same qualitative behavior for the MGTS task in Figs.~\ref{fig:tasks_vs_trotter}~(g) and (h).
There, however, the difference between the chaotic and integrable case is less drastic. 
For the NARMA task in Figs.~\ref{fig:tasks_vs_trotter}~(i) and (j), we see that, in analogy to the STM case, there is a crossover at $\tau = L$ where the DU gates perform \emph{worse}.
For smaller delays, there is now however a much wider range of values for the Trotter step which provide a good performance, including, but not limited to, the DU point.

Finally, we further stress that we have chosen $J_z = 0.8$ for these simulations.
As we discuss in Sec.~\ref{sec:benchmark_entanglement}, this corresponds to gates with relatively \emph{small} entangling power. 
When $J_z$ is chosen such that these gates provide large or maximum entangling power, some of these results can look quite different (see Appendix~\ref{app:trotter_high_entanglement}).
Even more importantly, in some of these scenarios the DU point might \emph{not} lead to the optimal configuration.
An extensive search over the full space of unitary gates (see Appendix~\ref{app:weyl}) indicates that the optimal performance can generally be found at or near the dual-unitary point, but dual-unitarity in itself does not guarantee an optimal performance.

\subsection{The role of entanglement}
\label{sec:benchmark_entanglement}

\begin{figure}
    \centering
    \includegraphics[width=\columnwidth]{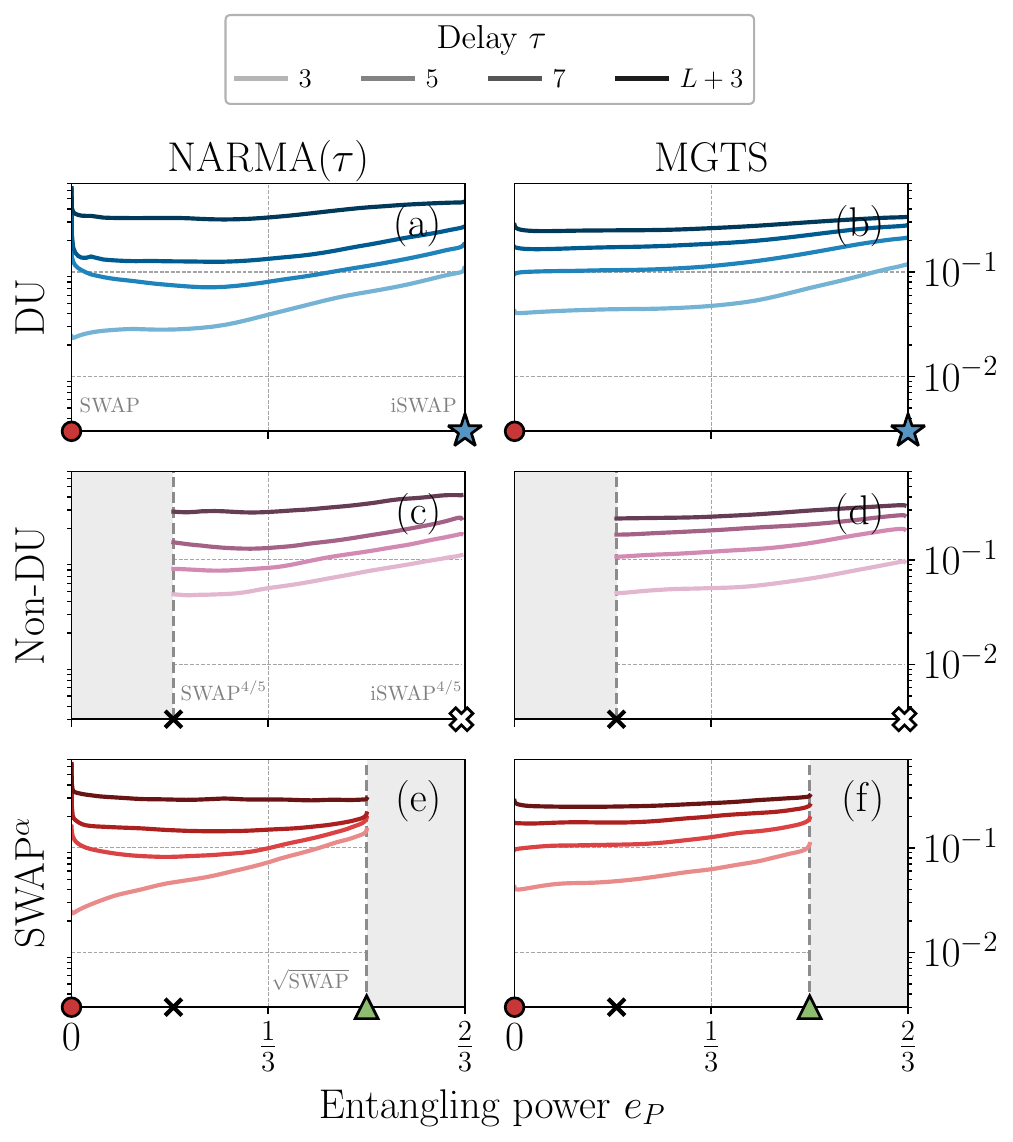}
    \caption{
    Task error plotted as $1 - C$ as a function of the entangling power~[Eq.~\eqref{eq:entangling_power}] (see App.~\ref{app:weyl_chamber_ep} for the general expression).
    We plot results for three types of gates: (a, b) dual-unitary (blue), (c, d) generic non-DU gate with $\Delta t = \pi/5$ (pink) and (e, f) the family of fractional SWAP gates (red).
    We consider the NARMA($\tau$) and MGTS tasks.
    The different curves in each panel correspond to the delay choice $\tau$, where darker lines are associated with larger delays.
    Each point is averaged over $2000$ realizations, corresponding to $50$ circuit and $40$ task realizations.
    We mark a few special points for each family of gates:
    SWAP gate (red circle), iSWAP (blue star), SWAP$^{4/5}$ (black cross), iSWAP$^{4/5}$ (white cross) and $\sqrt{\mathrm{SWAP}}$ (green triangle).
    In this figure we have chosen $L = 7$.
    }
    \label{fig:tasks_vs_entanglement}
\end{figure}

We now move forward and investigate how the entangling properties of the two-qubit gates used to build the Floquet unitary [Eq.~\eqref{eq:floquet_circuit_drawn}] affect the performance of the reservoir.
Our quantity of choice will be the entangling power of two-site gates~\cite{zanardiEntanglingPowerQuantum2000, zanardiEntanglementQuantumEvolutions2001}.
Given a two-qudit gate, the entangling power measures the entanglement generated by a gate $U$, on average, by its action upon product  states.
Mathematically:
\begin{equation}
    e_P(U)
    =
    C_q \, \mathbb{E}_{\mathrm{Haar}}
    \left[
    E(U \, \ket{\phi_1} \otimes \ket{\phi_2}
    \right],
\end{equation}
with the average $\mathbb{E}_{\mathrm{Haar}}[\bullet]$ being performed over the single-site states $\ket{\phi_1}$ and $\ket{\phi_2}$, independently drawn from the Haar measure. 
The prefactor $C_q$ denotes an unimportant normalization constant 
~\footnote{
Note that many previous works do not assume any particular normalization, and simply take $C_q = 1$.
In this case, the maximum value for the entangling power is $2/9$.
Here we choose $C_q$, such that the maximum entangling power is $2/3$, in order to be consistent with adjacent works on dual-unitary circuits.
}.
Finally, $E$ denotes an arbitrarily chosen measure of entropy.
When $E$ is chosen to be the linear entropy, the entangling power assumes a particularly simple form and
the Cartan decomposition for DU gates in Eq.~\eqref{eq:gate_parametrization} conveniently allow us to parametrize the entangling power of the gate in terms of the anisotropy $J_z$. 
Namely, it can be shown that the entangling power associated with the two-qubit gate from Eq.~\eqref{eq:gate_parametrization} is given by~\cite{balakrishnanEntanglingPowerLocal2010}:
\begin{equation}\label{eq:entangling_power}
    e_P
    =
    \frac{2}{3}\cos^2{\left(\frac{\pi}{2} J_z \right)},
\end{equation}
which attains its maximum at $J_z = 0$, corresponding to the iSWAP gate, and vanishes at $J_z = 1$, corresponding to the SWAP gate.
We supplement the general formula, for arbitrary Trotter step $\Delta t$, in App.~\ref{app:weyl_chamber_ep}.
This works as an interesting resource because the entangling power is invariant under transformations which depend only on local unitaries~\cite{ratherCreatingEnsemblesDual2020}.
Thus, although changing the single-qubit operators $u_+$ and $u_-$ in Eq.~\eqref{eq:gate_parametrization} can lead to a broad range of behaviors when correlation functions and integrability/ergodicity are considered, the entangling power $e_P$ itself remains the same.
For that reason, by changing only the anisotropy parameter $J_z$ and leaving everything else fixed, we can tune the entangling power of the two-qubit unitaries in the circuit and get some intuition on its role on the performance of the reservoir.

Basic results are shown in Fig.~\ref{fig:tasks_vs_entanglement}, where we plot the error $1-C$ as a function of the entangling power.
We separate the simulations as follows: 
the left column, with panels (a, c, e), correspond to the NARMA task, while the right column with panels (b, d, f) corresponds to the MGTS task.
Each row corresponds to a different type of circuit.
In the first row [panels~(a, b)], we plot the results for the Floquet circuit built from chaotic dual-unitary circuits.
For the second row [panels~(c, d)], we consider a family of non-DU circuits instead, where the Trotter step is chosen to be $\Delta t = \pi/5$.
For completeness we also consider the family of fractional SWAP circuits [panels~(e, f)].
This allows us to interpolate within this family, from the non-entangling to the entangling $\sqrt{\mathrm{SWAP}}$ gate.
We discuss these in more detail in App.~\ref{app:weyl_chamber_ep}.
For the first row, the entangling power is computed through Eq.~\eqref{eq:entangling_power}, while for the latter two we use the formula given by Eq.~\eqref{eq:entangling_power_general} in App.~\ref{app:weyl_chamber_ep}.

Surprisingly, increasing the entangling power too much is often detrimental. 
We see that, for most of the tasks and delays, an intermediate amount of entanglement is optimal (depending on the specific value of $\tau$ we consider).
For sufficiently large memory requirements, the operation is optimal for slightly perturbed systems when we consider the MGTS task, where $e_P$ is somewhat small.
For some tasks however, the SWAP gate (with $e_P = 0$) leads to optimal performance (within that a given family).
This is seen in the NARMA task for the DU and fractional SWAP families at $\tau = 3$, where the non-entangling gate wins.
Analogous to the arguments in Sec.~\ref{sec:role_soliton_feature}, these results are a direct consequence of our reservoir architecture and the properties of the SWAP gate:
the SWAP circuit trivially recovers the inputs up to $\tau = L - 1$, so pure (short) memory tasks are trivial in this case.
However, in counterpoint to what happens to the SWAP circuit case, the entangling circuits do not suffer under the limitation of failing to retain memory for large $\tau$. 
In panel (e), for example, we see that as long as we consider larger delays, the curves display a sharp jump in the performance as soon as $e_P > 0$.
The reason is that both nonlinear effects and information about inputs further away in the past are better spread out through the reservoir and can be recovered later on (as sketched in Sec.~\ref{app:operator_growth}).
Overall, these results indicate that fine-tuned circuits (such as the SWAP case), can lead to good results for very specific tasks (STM, parity check) as long as the observables/features are appropriately chosen.
Nevertheless, as soon as the chosen features are more arbitrarily chosen, and more complex tasks are considered (NARMA, MGTS), then having features which make the dynamics less trivial and the reservoir more expressive, such as entanglement, can lead to better, less fine-tuned and more agnostic results.

Recent investigations for other setups, such as in Ref.~\cite{askariSpinNetworkQuantumReservoir2025}, have led to similar numerical results.
There, the authors compute the entanglement between different bipartitions in a spin model and analyse the corresponding performance.
Depending on how entanglement is spread among these qubits, the QRC operates in different ways.
In particular, they also find that a moderate amount of entanglement typically leads to optimality.
Another instance is Ref.~\cite{karimiRoleEntanglementQuantum2025}, which considers coupled Kerr oscillators.
There, entanglement also generally leads to enhancement, as long as one is careful with the choice of parameters in the model.
In both cases, the message is consistent with our results.


\section{The role of features}
\label{sec:features}

After having discussed the choice of gates and the role of entanglement, let us now consider the role of the choice of features. For ergodic dynamics the results are largely independent of the choice of features, whereas in the integrable case the performance of the reservoir depends strongly on this choice, such that we here focus on integrable circuits.
Supplementary investigations for other choices of features, including in the ergodic case,  are given in Appendix~\ref{app:choice_features}.

\subsection{Memory through solitons}
\label{sec:role_soliton_feature}

In this section we justify the results seen in the previous section pertaining to integrable circuits. 
We explain the role of the solitons in the erase-input dynamics and the memory effects in the reservoir.
We initially consider the observables $O_j$ to be single-body solitons at arbitrary sites in the chain and assume odd circuit size $L$. 
In a slight abuse of notation, we reuse the same graphical notation as the one introduced in Eq.~\eqref{eq:soliton_graphical_definition} for the folded picture, which should be unambiguous from the context.

Let us suppose, without loss of generality, that the soliton $Z_{\tau_s}$ is placed at an odd site $2\tau_s + 3$, with $\tau_s \in \mathbb{N}$ being a chosen delay value.
The expected value of this observable, in the folded picture, is:
\begin{equation}\label{eq:soliton_identity_folded}
    \mathrm{tr}\{\rho_t Z_{2\tau_s + 3}\}
    =
    \includegraphics[valign=c]{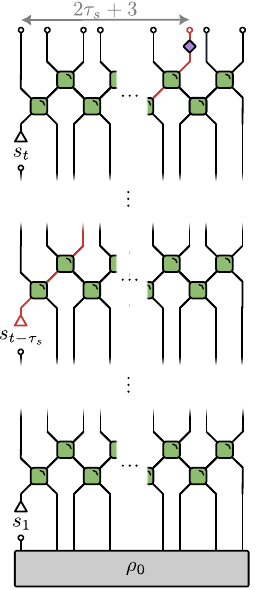},
\end{equation}
where the unimportant initial state is given by $\rho_0$.
Following the soliton identity in the folded picture, as depicted in Eq.~\eqref{eq:soliton_folded}, we see that the chosen soliton is shifted to the left.
The corresponding path is highlighted by the red wire in the equation above.
Due to the erase-input part of the dynamics, the left-moving soliton will eventually encounter an input state $\ket{s_{t - \tau_s}}$ at earlier time $t - \tau_s$, as we equivalently show in:
\begin{equation}
    \mathrm{tr}\{\rho_t Z_{2\tau_s + 3}\}
    =
    \includegraphics[valign=c]{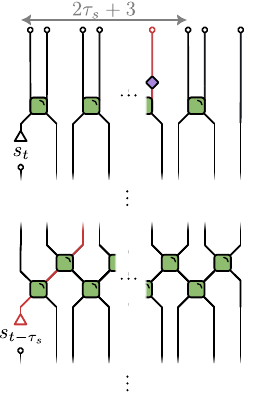}.
\end{equation}
For visual clarity we have omitted the initial state and applied the unitarity condition to the topmost layer only.
We can now use unitarity of the gates and the soliton property, as depicted in Eqs.~\eqref{eq:folded_unitarity} and~\eqref{eq:soliton_folded}, respectively, to repeatedly apply the simplification above and eliminate all unitaries from the equation.
Hence, for solitons placed at odd sites, we eventually have:
\begin{equation}\label{eq:soliton_feature_result}
    \mathrm{tr}\{\rho_t Z_{2\tau_s + 3}\}
    =
    \includegraphics[valign=c]{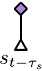}
    =
    \bra{s_{t - \tau_s}}Z\ket{s_{t - \tau_s}}
    =
    2s_{t - \tau_s} - 1.
\end{equation}
%

\begin{figure}
    \centering
    \includegraphics[width=0.95\columnwidth]{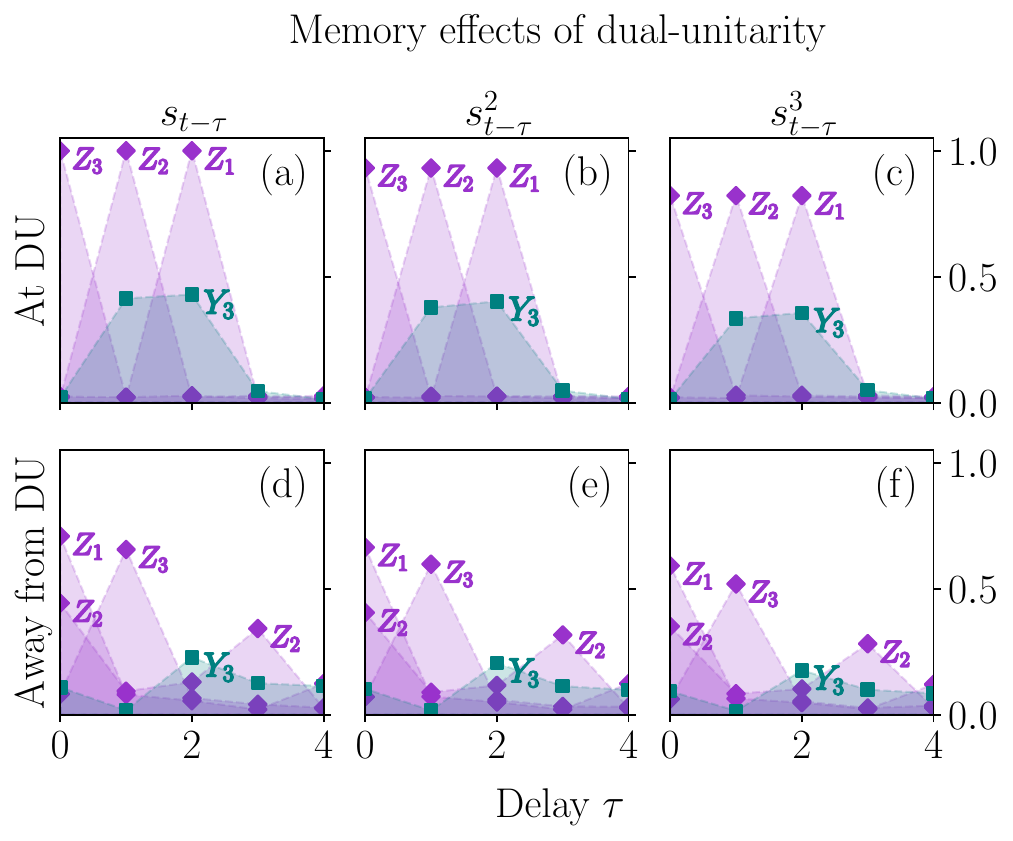}
    \caption{
    Plot of the IPC associated with single features. 
    Each curve corresponds to a training-validation run where only a single feature was used.
    We increase the degree of nonlinearity for the STM task along the columns in the plot.
    Panels (a)-(c) in the top row correspond to the dual-unitary circuit, where we highlight, in purple, the three single-body solitons $Z_1, Z_2, Z_3$ (purple diamonds).
    Likewise, we also plot the performance for single-body Pauli operators $Y_i$ (teal squares).
    Panels (d)-(f) in the bottom row show the corresponding results away from dual unitarity, for a non-DU configuration with $\Delta t \neq \pi/4$.
    }
    \label{fig:tasks_vs_feature_DU}
\end{figure}

By extending the same argument to even sites and the edges, we obtain the generalization:
\begin{equation}\label{eq:soliton_as_feature}
    2 s_{t - \tau_s}
    =
    \begin{cases}
        1 + \expec{Z_{2\tau_s + 3}}&, \: \tau_s < L/2 - 1 \\
        1 + \expec{Z_{2(L - \tau_s - 1)}}    &, \: L/2 - 1 < \tau_s < L - 1\\
        1 + \expec{Z_1}            &, \: \tau_s = L - 1
    \end{cases}
\end{equation}
In other words, there is a one-to-one correspondence between the time-delayed input $s_{t-\tau_s}$ and the expected value of the solitons.
We can also see that it is not possible to retrieve information about the inputs for $\tau \geq L$ by employing the solitons: eventually, and at most with delay of $L - 1$, all solitons will meet one of the input states.
In this case, it is necessary to consider more generic operators, which grow and spread over the lattice in order to properly capture information at later times.
The intuition behind this can also be understood from the discussion in Sec.~\ref{app:operator_growth}.  
These observations constitute an explanation of why we obtain perfect accuracy in the STM task for small delays, as discussed in regard to Fig.~\ref{fig:tasks_vs_delay}~(a).
Note that the end effect of using these solitons is similar to the memory augmentation achieved via classical post-processing in Ref.~\cite{settinoMemoryAugmentedHybridQuantum2024}.
In our case, however, this feature is embedded into the circuit structure itself, due to the presence of solitons.

\begin{figure}
    \centering
    \includegraphics[width=0.95\columnwidth]{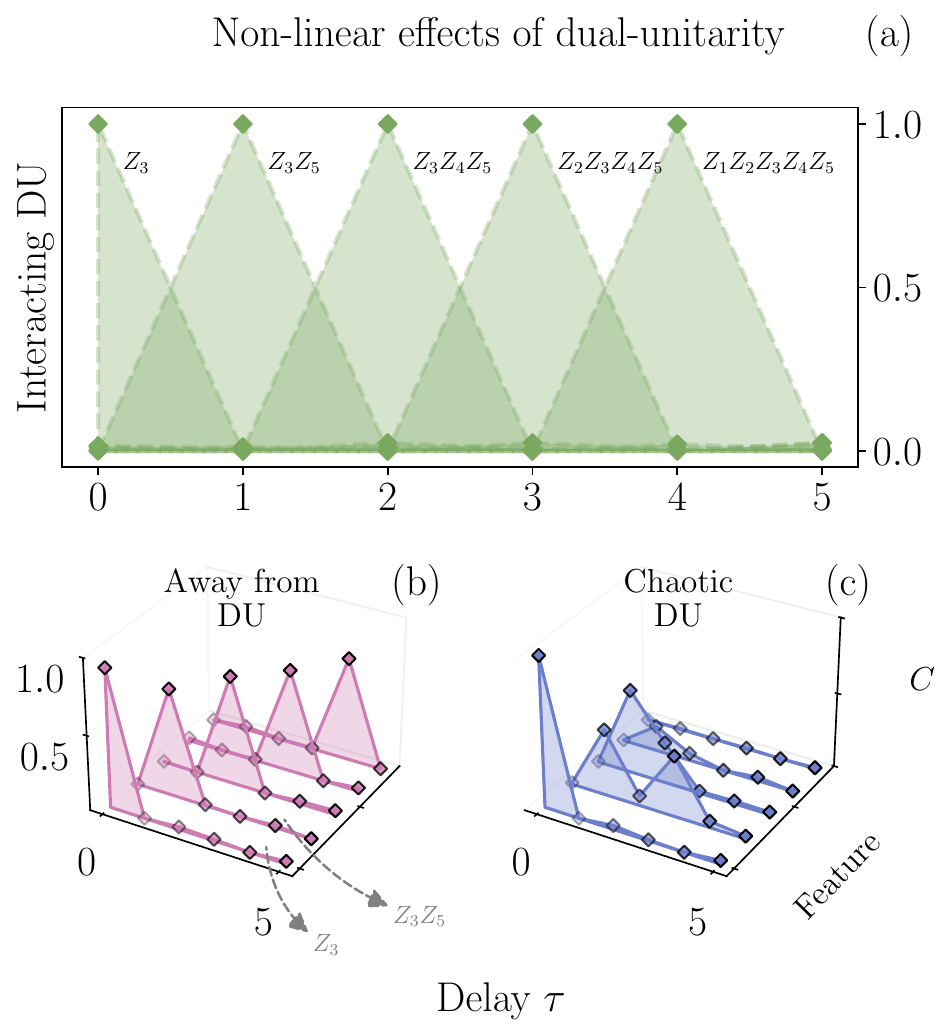}
    \caption{
    Contribution of the selected $n$-body $Z$-type features to the IPC for the parity task. 
    (a) Interacting integrable circuit at the dual-unitary point.
    (b) Integrable circuit away from the dual-unitarity with $\Delta t = \pi/3$.
    (c) Chaotic circuit at the dual-unitary point.
    In the 3D plots, each curve corresponds to the IPC as a function of the delay $\tau$, for each of the features highlighted in panel (a).
    In the dual-unitary integrable case, each soliton contributes to the memory for a single delay only.
    In the chaotic case, the memory effects spread out across the lattice, and the Pauli operators $Z$ can recover (correlated) information in a more generic manner.
    }
    \label{fig:tasks_vs_feature_DU_parity_3D}
\end{figure}

These properties are further illustrated with numerics in Fig.~\ref{fig:tasks_vs_feature_DU} for circuit size $L = 3$.
We focus on the first row [panels (a)-(c)], where simulations were performed for the interacting integrable circuit ($\epsilon = 0$), corresponding to the Trotterized XXZ circuit (defined in Eq.~\eqref{eq:circuit_int} and associated with the green curves in Figs.~\ref{fig:tasks_vs_delay} and~\ref{fig:tasks_vs_trotter}).
In this figure, the training is performed on the basis of a \emph{single} feature.
This approach allows us to capture and \emph{isolate} their individual contributions to the processing capacity of the reservoir.
First, we plot and highlight the learning capacity according to single-body solitons in purple.
We can see that the soliton $Z_3$ retrieves the input with no delay, the soliton $Z_2$ retrieves the input with delay $\tau = 1$ and the soliton $Z_1$ retrieves the input for delay $\tau = 2$, agreeing with the analytical statement in Eq.~\eqref{eq:soliton_as_feature}.
This can be seen through the peaks in the figure.
Adding nonlinearity in the STM decreases the performance but maintains the same peak structure, since the single-solitons themselves can only capture linear memory contributions, but not nonlinear effects.

Another aspect of this plot is the behavior of more generic observables, besides solitons. By the previous argument, the solitons are associated with a specific input $s_{t - \tau_s}$, and that input only.
The Pauli matrix $Y$ (teal curve), on the other hand, is able to capture information about many different time-lags, including those for which $\tau_s \geq L$.
Hence, precisely because of the operator growth it undergoes (as we discuss in Sec.~\ref{app:operator_growth}), this feature is able to retrieve information in the circuit for many different time steps.
In the bottom row [panels (d)-(f)] we display the results for the more generic non-DU configuration ($\Delta t \neq \pi/4$).
As such circuits do not have any solitons associated with them, we lose the previous properties.
Instead, we can now see that the $Z$ operators start to behave more generically, and they are not restricted to recovering information at a single time step any longer -- they can recover information at different times (as seen with $Z_2$, for instance).
Nevertheless, the overall capacity of the features seems to be diminished, both for the $Y$ and $Z$ Pauli matrices.

To conclude, we highlight how in the integrable case the solitons and these features can capture nonlinear effects.
It is straightforward to check that solitons do not interact, i.e. $ V^\dagger (\soliton \soliton) V =  (\soliton \soliton)$, which means that neighboring solitons pass each other by.
Let us consider, without any loss of generality, two delays $\tau_s$ and $\tau_s'$, such that $\tau_s < \tau_s' < L/2 - 1$.
In analogy to the derivation for Eq.~\eqref{eq:soliton_feature_result}, we can show, for example, that
\begin{equation}\label{eq:soliton_parity}
\begin{split}
    \mathrm{tr}\{\rho_t Z_{2\tau_s + 3}Z_{2\tau_s'+ 3}\}
    &=
    \includegraphics[valign=c]{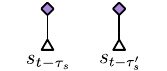}\\
    &=
    \expec{Z_{2\tau_s + 3}}
    \expec{Z_{2\tau_s' + 3}}\\
    &=
    (2s_{t - \tau_s} - 1)(2s_{t - \tau_s'} - 1).
\end{split}
\end{equation}
The expected value of non-local solitons simply factorizes in terms of products of local terms, and these will ultimately yield multi-time correlations/nonlinearities between the inputs.
For instance, consider the parity check task~\eqref{eq:parity_task} with $\tau=1$. 
We have that $y_t = (s_t + s_{t-1}) \mod 2$.
By considering the observable $O_j = Z_3 Z_5$ and $L \geq 5$ we get 
$x_{t, j} = \bra{s_{t}}Z\ket{s_{t}}\bra{s_{t - 1}}Z\ket{s_{t - 1}}
= (2s_t - 1)(2s_{t-1} - 1)$,
which is equivalent to $1 - 2y_t$ and therefore is linearly related to the target function $y_t$.
This is shown in Fig.~\ref{fig:tasks_vs_feature_DU_parity_3D}.
Therefore, the challenge when solitons are chosen as features for this task arises due to the fact that large memories in the parity check require highly non-local observables.
For instance, in order to fully reproduce the parity check for $\tau = L - 1$, one needs to consider the full product $Z_1 Z_2 ... Z_L$, with support over the whole circuit.

This is further illustrated in Fig.~\ref{fig:tasks_vs_feature_DU_parity_3D} by plotting the contributions of five selected features to the parity check task when they are taken as the sole features, up to $\tau = 4$. These operators are chosen from the $Z$-type observables that act as solitons in the dual-unitary integrable circuit.
We show the results for the interacting integrable circuit [Fig.~\ref{fig:tasks_vs_feature_DU_parity_3D}~(a), green plot], integrable circuit but non-DU [Fig.~\ref{fig:tasks_vs_feature_DU_parity_3D}~(b), pink plot] and for a generic ergodic circuit [Fig.~\ref{fig:tasks_vs_feature_DU_parity_3D}~(c), blue plot].
In panel~(a) we see that, just like in the STM case, we associate the learning for a given delay $\tau$ with a given (multi-body) soliton.
For instance, in order to learn the parity check task for $\tau = 2$, corresponding to the middle peak in panel~(a), we need to consider the feature $Z_3 Z_4 Z_5$.
In panels~(b) and~(c) we plot a 3D version of this plot, where we show the IPC as a function of the delay $\tau$. 
The different curves correspond to the five chosen features, as displayed in panel~(a). 
In the integrable case away from the DU point, we see a similar qualitative behavior: when these features behave as solitons in the dual-unitary integrable circuit, there is a one-to-one correspondence between them and the delay, albeit the information recovery is not perfect anymore and the IPC is thus smaller than unity.
In panel~(c) we see a more generic behavior, where now each of the operators will non-uniquely capture correlations.
For instance, for the second curve, from bottom to top, we see that the feature $Z_3 Z_5$ displays two peaks at delays $\tau = 1$ and $\tau = 3$ for the parity task.
The peak at $\tau = 1$ means that $Z_3 Z_5$ learns about the correlations between $s_{t}$ and $s_{t-1}$, while the peak at $\tau = 3$ corresponds to correlations between $s_{t}, \dots, s_{t-3}$. 
Hence, while capturing correlations up to $s_{t-3}$ required a \emph{four}-body operator in the integrable case, in the ergodic regime this could be achieved with a simpler two-body operator.

\subsection{Memory in ergodic circuits}
\label{sec:memory_ergodic}
In the ergodic case, there is no longer a notion of solitons. Still, for sufficiently short delays, we see in Fig.~\ref{fig:tasks_vs_delay} that the performance of the reservoir remains qualitatively similar. 
The corresponding mechanism is similar: In dual-unitary circuits quantum information gets perfectly transported along the edge of the so-called `causal light cone', i.e. along the path of the solitons in the integrable case, allowing for perfect decoding in which the input can be perfectly reconstructed from the output provided the unitary evolution operator is known~\cite{RamppHP2024}. 
As such, if the unitary evolution operator is known, the input over the past $\tau$ time steps can be perfectly reconstructed from the corresponding single-site features on the $\tau$ first sites.
The key difference is that in the integrable case this recovery does not require knowledge of the unitary evolution operator since it acts trivially on the solitons (features).
This perfect recovery is closely related to the fact that in dual-unitary circuits nonvanishing single-site features can only originate from their light cone, i.e. from the corresponding (delayed) input~\cite{BertiniExact2019}.
The imperfect memory in the ergodic dual-unitary case is hence due to the imperfect learning of the unitary evolution operator.
In generic quantum circuits such a decoding necessarily induces an additional error and no perfect recovery is possible even if the unitary evolution operator is fully known.
Combined, these explain why the memory effects of this operator dynamics extend beyond the integrable case to the ergodic case, as apparent from Fig.~\ref{fig:tasks_vs_delay}.

\subsection{Nonlinear processing through operator growth}
\label{app:operator_growth}

In this section we illustrate how single-site local observables can recover memory and nonlinear effects in the dual-unitary integrable XXZ circuit. While we focus on the integrable case for concreteness, these results again directly extend to generic circuits.
We start by noting that the controlled-phase gate~\eqref{eq:cphase_gate} can be rewritten, in terms of Pauli matrices, as:
\begin{equation}
    \CPhase
     \propto
    \exp
    \left[
        \frac{i \phi}{4} (Z \otimes I + I \otimes Z + Z \otimes Z)
    \right] 
    ,
\end{equation}
up to an unimportant global phase.
We introduce some supplementary notation to Eq.~\eqref{eq:soliton_graphical_definition}, where we also define the remaining Pauli operators graphically:
\begin{equation}\label{eq:X_and_Y_graphical}
    X_x
    =
    \hdots
    \underset{x - 1}{\leg}
    \:
    \underset{x}{\Xdraw}
    \:
    \underset{x + 1}{\leg}
    \hdots,
    \quad
    Y_x
    =
    \hdots
    \underset{x - 1}{\leg}
    \:
    \underset{x}{\Ydraw}
    \:
    \underset{x + 1}{\leg}
    \hdots,
\end{equation}
The action of this gate on the $Y$ operator, given by the conjugation $V^\dagger (Y \otimes \textbf{1}) V$, reads (in \emph{folded} notation):
\begin{align}\label{eq:conjugate_YI}
    \includegraphics[valign=c]{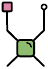}  
    =\,
    -&\frac{1}{2}\sin \phi \,
    \reflectbox{\includegraphics[valign=c]{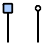}}  
    +
    \left( \frac{1 + \cos \phi}{2} \right) \,
    \reflectbox{\includegraphics[valign=c]{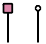}}  \nonumber\\
    -&\frac{1}{2}\sin \phi \,
    \reflectbox{\includegraphics[valign=c]{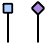}}  
    -
    \left( \frac{1 - \cos \phi}{2} \right)\,
    \reflectbox{\includegraphics[valign=c]{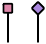}}  
    \,.    
\end{align}
After this local conjugation, the first two contributions on the RHS correspond to purely local terms, while the last two already have two-site support and can seed further operator growth in subsequent layers.

Now the question is: does this provide any intuition on how exactly the circuit encodes the input information?
As a concrete example, we take the observable $Y_2$ for a chain of arbitrary size $L$.
We then use the two identities above to investigate explicitly how this feature behaves under the circuit dynamics. 
In order to do so, let us consider a sequence of inputs up to $\ldots, s_{t-1}, s_t$ at time $t$.
After applying Eq.~\eqref{eq:conjugate_YI} 
to compute its expected value, we get:
\begin{widetext}
\begin{align}
    \label{eq:conjugated_Y_circuit}
    \tr{\rho_t Y_2}
    &=
    \includegraphics[valign=c]{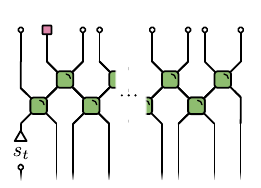}\\  
    &=
    \label{eq:conjugated_Y_circuit_reduced_A}
    - \frac{1}{2}\sin \phi \:
    \includegraphics[valign=c]{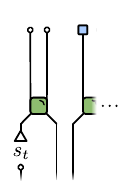}
    +
    \left( \frac{1 + \cos \phi}{2} \right) \,
    \includegraphics[valign=c]{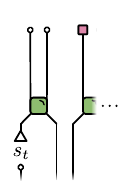}
    -\frac{1}{2}\sin \phi \:
    \includegraphics[valign=c]{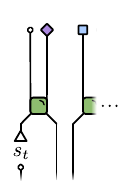}
    -
    \left( \frac{1 - \cos \phi}{2} \right) \:
    \includegraphics[valign=c]{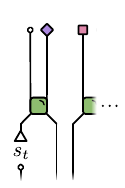}\\ 
    &=
    \label{eq:conjugated_Y_circuit_reduced_B}
    -\frac{1}{2}\sin \phi \:
    \includegraphics[valign=c]{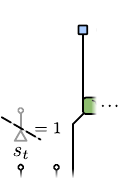}
    +
    \left( \frac{1 + \cos \phi}{2} \right) \:
    \includegraphics[valign=c]{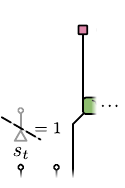}
    -\frac{1}{2}\sin \phi \:
    \includegraphics[valign=c]{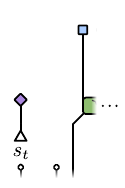}
    -
    \left( \frac{1 - \cos \phi}{2} \right) \:
    \includegraphics[valign=c]{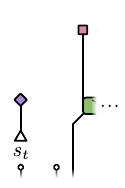}\\ 
    &=
    \label{eq:conjugated_Y_circuit_reduced_C}
    - s_t \sin \phi \:
    \includegraphics[valign=c]{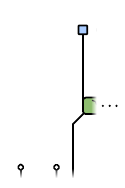}
    +
    \left[
    1 - s_t + s_t \cos{\phi}
    \right]
    \:
    \includegraphics[valign=c]{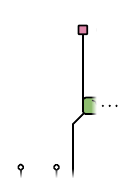}
    \,,   
\end{align}
\end{widetext}
where we have omitted most of the circuits and all the inputs before $s_t$ for clarity.
In the last two lines we have used the fact that $\bra{s_t}Z\ket{s_t}$ evaluates to $2s_t - 1$, as discussed in Sec.~\ref{sec:role_soliton_feature}.
Hence, just like the solitons, other Pauli operators can also encode individual inputs.
Now, to see how we recover time-correlated contributions, let us consider the second circuit contribution on the RHS of Eq.~\eqref{eq:conjugated_Y_circuit_reduced_C}, for concreteness.
If we isolate one contribution with a single $Y$ operator, the full expansion contains a term of the type:
\begin{widetext}
\begin{equation}
    \left[
    1 - s_t  + s_t \cos{\phi}
    \right]
    \:
    \includegraphics[valign=c]{images/conjugated_Y_circuit_4_a.pdf}
    \propto
    s_t
    \includegraphics[valign=c]{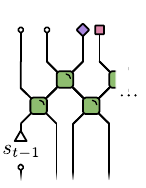}
    +
    \cdots
    =
    s_t
    \Tr{\rho_{t - 1} Z_3 Y_4}
    +
    \cdots
    \propto
    s_t s_{t-1} 
    \:
    \includegraphics[valign=c]{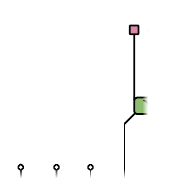}
    +
    \cdots ,
\end{equation}
\end{widetext}
which we have further expanded reapplying the identity from Eq.~\eqref{eq:conjugate_YI} and the soliton resolution.
This is analogous to the simplification made in the last term of Eqs.~\eqref{eq:conjugated_Y_circuit_reduced_A} and~\eqref{eq:conjugated_Y_circuit_reduced_B} to the $Z \otimes Y$ contribution.
Note that the proportionality in the equation above hides phase-dependent prefactors
and the ellipses denote the remaining Pauli-string contributions.
This derivation sketches how local observables can contain temporally correlated inputs through the factors $s_t s_{t - 1}$ or analogous combinations.

This result depicts how operator growth gives rise to nonlinearity.
While soliton dynamics is restricted to integrable circuits, dual-unitary circuits generally exhibit a maximal operator growth~\cite{bertiniOperatorEntanglementLocal2020,claeysMaximumVelocityQuantum2020,BertiniScrambling2020}, and the above results can be directly generalized to generic dual-unitary circuits.
Taken together, these results indicate that the maximal operator growth of dual-unitary circuits underlies their optimal performance in certain tasks.

\section{Repeated-input scheme for nonlinearity}
\label{sec:nonlinearity_legendre}

We now investigate how to choose the features and recover same-time nonlinearities.
As we have observed in Fig.~\ref{fig:tasks_vs_feature_DU}~(b) and~(c), even local operators, or single-Pauli solitons in the integrable case, can learn the nonlinear STM task ($d = 2, 3$) reasonably well.
However, this numerical result is somewhat misleading: 
the linear task $s_{t - \tau}$ displays a very high correlation with its nonlinear counterpart.
Indeed, assuming $s_t$ is sampled from the uniform distribution $s_t \sim U(0, 1)$ and that $s_t$ is perfectly learnable, one would find $C \approx 0.94$ between $s_{t-\tau}^2$ and $s_{t-\tau}$.
Thus, the ``good" performance we observe in Fig.~\ref{fig:tasks_vs_feature_DU} is an artifact of the favorable overlap between both versions of the STM task.
This has been extensively explored in the classical RC literature, such as in Ref.~\cite{dambreInformationProcessingCapacity2012}, where a fundamental trade-off between memory and nonlinearity has been established.
See also Ref.~\cite{llodraBenchmarkingRoleParticle2023} for this benchmark in a proper quantum reservoir.
For that reason, in this section we will consider the family of Legendre polynomials of order $\ell$, taking $2s_t - 1$ as an argument.
We will denote them by $P_\ell(2s_t - 1)$. 
Polynomials of distinct degree will have vanishing correlation [Eq.~\eqref{eq:capacity_definition}], by construction.
This approach will allow us to separate truly nonlinear effects from basic input memorization, and better pinpoint the role of different properties of the reservoir. 

A limitation of the usual erase-input protocol in a digital approach like ours is that we \emph{cannot} encode arbitrary polynomial functions of $s_t$.
More specifically, by writing down the density matrix of the encoding state in Eq.~\eqref{eq:erase_input}, we see that the dependence on a single input is restricted to the space $\mathrm{span}(1, s_t, \sqrt{s_t (1 - s_t)})$.
A way to build intuition for this is through the calculations shown in Sec.~\ref{app:operator_growth}.
In essence, operator growth or the use of non-local features (see Sec.~\ref{sec:role_soliton_feature}) only helps with temporal correlations, that is, nonlinearities involving distinct inputs.
To circumvent this problem, we consider a \emph{repeated-input} strategy.
In that case, the same value is injected for $q$ consecutive steps before moving to the next sample in the sequence.
All the equations and the formalism from Sec.~\ref{sec:QRC} still apply.
A straightforward way to implement this strategy is to take the original input vector $\boldsymbol{s}$ and transform it into a new vector $\boldsymbol{s}^{(q)}$, where each entry is repeated $q$ times.
In our case, we focus on $q = 3$ to illustrate the approach, such that, given the original vector $\boldsymbol{s} = (s_1, s_2, \dots)$, we now have $\boldsymbol{s}^{(3)} = (s_1, s_1, s_1, s_2, s_2, s_2, \dots)$.
This encoding strategy allows correlations involving the same input $s_t$ to build up, yielding powers of $s_t$ up to order $q$.
As we show in our numerics, the trade-off is that part of the available memory window is spent re-processing the same input, so long-delay performance is expected to deteriorate.
This approach shifts the memory cross-over, or the sharp cut-off in some of the integrable cases, to shorter delays.
In the numerics shown below, this crossover appears to move approximately as $L - \ell$, with higher-order Legendre tasks losing support at earlier delays.

\begin{figure*}
    \centering
    \includegraphics[scale=1]{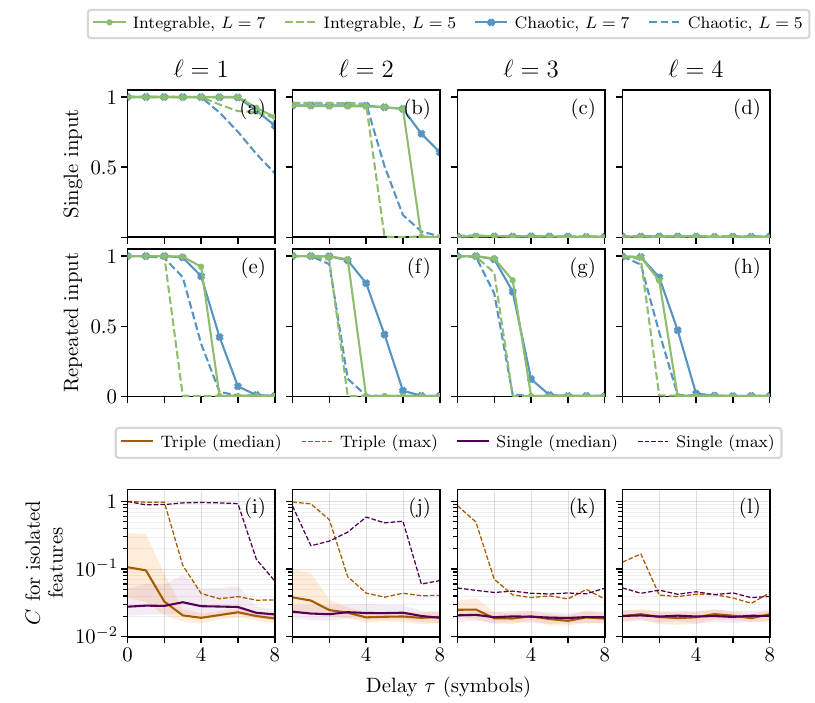}
    \caption{
    Performance of the reservoir for orthogonal nonlinear tasks based on Legendre polynomials of different orders $\ell$ (columns).
    Results are shown for integrable (green circular markers) and chaotic circuits (blue crosses), with dashed and solid lines corresponding to $L = 5$ and $L = 7$, respectively.
    IPC for the single-input scheme in panels (a) - (d), and for the repeated-input scheme (three repetitions) in panels (e) - (h).
    In the bottom row, panels (i) - (l) show the IPC for single-feature training for the single-input (purple) and repeated-input (orange) schemes in the $L = 7$ chaotic circuit.
    The central line represents the median, the upper line the maximum, and the shaded area the quartiles.
    We consider a distribution of the information processing capacity for $1155$ features (Pauli strings of up to three operators).
    The first two rows indicate the trade-off between memory and nonlinear effects:
    by introducing the repeated-input scheme, we sacrifice long-term memory for higher nonlinear capabilities.
    Note how the single-input scheme is unable to learn the third- and fourth-order Legendre polynomials, $\ell = 3, 4$.
    The last row shows how individual features contribute to the learning of the Legendre polynomials.
    For the first two rows, simulations were run for $350L$ and $50L$ training and validation steps, respectively. 
    In the case of the isolated runs at the last row, a fixed number of $250$ training points was used.
    }
    \label{fig:features_legendre}
\end{figure*}

The results are shown in Fig.~\ref{fig:features_legendre}.
In this new scenario, we consider a much larger number of nodes in the training.
We take \emph{all} Pauli strings up to weight $3$.
For $L = 7$, that corresponds to $1155$ features.
In the first row [Figs.~\ref{fig:features_legendre}~(a) - (d)], we show the results for the single-input strategy.
In the second row [Figs.~\ref{fig:features_legendre}~(e) - (h)], we consider the repeated-input strategy.
Finally, in the last row [Figs.~\ref{fig:features_legendre}~(i) - (l)], we show the contribution of isolated features for the $L = 7$ chaotic circuit.
In the first row, we plot the resulting IPC for each Legendre polynomial, up to order $\ell = 4$.
The first thing we observe is that, at both first and second order, there is a pronounced crossover around the system size.
This is particularly sharp for $L = 7$, corresponding to the solid lines.
At second order, we see that the chaotic reservoir displays slightly better performance for both system sizes.
As we have argued in a few other scenarios, the reservoir seems to better preserve nonlinearities in the chaotic regime, whereas the integrable reservoir might require more fine-tuning in the choice of features.
Finally, note that the learning completely breaks down at third and fourth orders, as we see in Figs.~\ref{fig:features_legendre}~(c) - (d).
Strikingly, Figs.~\ref{fig:features_legendre}~(g) and~(h) in the second row clearly illustrate how, by sacrificing some memory at the first two orders, we can increase the nonlinear capacity of the reservoir.
In this case, the integrable and chaotic reservoirs display very similar performance, with the latter being slightly better at the tails, and also for $\ell = 4$.
In Ref.~\cite{llodraBenchmarkingRoleParticle2023}, the authors find similar trade-offs.
In their case, however, the focus of the comparison lies on fermionic, bosonic and qubit-based systems.

Finally, we investigate how much \emph{individual} features contribute to this effect.
In Figs.~\ref{fig:features_legendre}~(i) - (l), corresponding to the last row, we repeat the same numerical analysis, but training is performed on the basis of a \emph{single} feature, in the same spirit as in Sec.~\ref{sec:role_soliton_feature}.
We perform this for all Pauli strings up to weight $3$ considered here.
Since highlighting the role of individual features would be difficult and not particularly illuminating, we focus instead on their \emph{distribution}.
In this row, we plot the IPC associated with all $1155$ features as a function of the delay $\tau$, highlighting the median, the maximum and the quartiles of that distribution.
In all four panels, we see that there is always (at least) one outlier feature with much larger IPC than the typical values in the distribution and the IPC eventually starts to diminish at all levels (median, maximum and quartiles).
This behavior closely follows the cut-offs observed in the first two rows.
We see the same trade-off here:
by switching from the single- to the repeated-input strategy, we decrease the IPC of the median and maximum at later times, but improve the results at third and fourth order.
Surprisingly, at third and fourth order, most features seem to hover around the noise floor, approximately $0.02$, with some small peaks in the quartiles and especially in the maximum-value curve.
This seems to suggest that, as long as single features are considered, very few of them can provide a significant response to the signals in this task.
Thus, we either need to cherry-pick a few very good features, which is however not an appropriate strategy for general scenarios, or greatly increase their number in order to better capture these nonlinearities.

\section{Finite-shot noise and exponential concentration}\label{sec:exp_concentration}

In this section we discuss how DU circuits fare against the problem of exponential concentration under the scenarios we have considered so far.
When we consider quantum approaches to machine learning, one would normally be led to believe that increasing the size of a reservoir likely leads to increased expressivity and improved performance.
This strategy, however, may actually lead to extra experimental hurdles in concrete scenarios~\cite{schutteExpressivityLimitsQuantum2025}.
One such challenge is the notion of \emph{exponential concentration}, which has recently been brought to the attention of the quantum machine learning community. 
A mathematical definition can be found, for instance, in Ref.~\cite{thanasilpExponentialConcentrationQuantum2024}.
What happens in practice is that, in sufficiently scrambling or random quantum models, increasing the Hilbert space dimension can lead to concentration of measure: as the number of qubits in the model increases, quantities of interest, such as expectation values, concentrate around a given (data-independent) value.
When this suppression becomes exponential in the number of qubits, one speaks of exponential concentration, hence the name.
Effectively, this means that a proper implementation may require an exponential number of measurement shots in order to achieve sufficient resolution -- otherwise what we obtain is, in practice, data-independent noise.
In such a situation the model becomes effectively input-blind and the reservoir approach is spoiled~\cite{thanasilpExponentialConcentrationQuantum2024, xiongFundamentalAspectsQuantum2025}.

\begin{figure}
    \centering
    \includegraphics[width=\columnwidth]{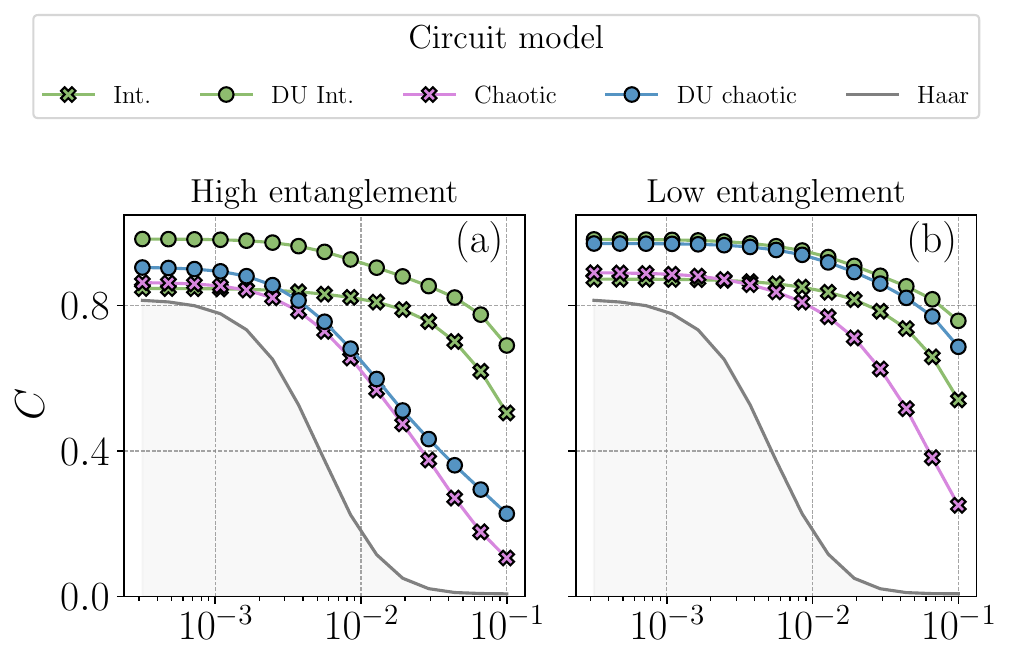}
    \includegraphics[width=\columnwidth]{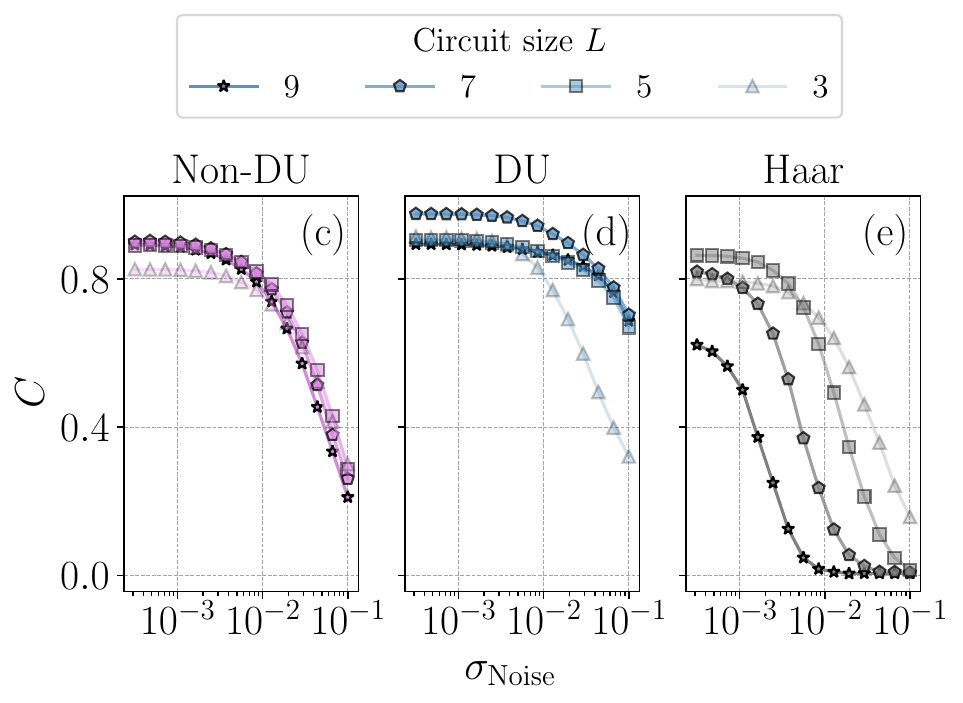}
    \caption{
    Effect of exponential concentration in the training of the brickwork circuit model.
    We plot the IPC for the NARMA($3$) task as a function of the finite-shot noise in two scenarios: (a) circuits with high entangling power ($J_z = 0.1$, corresponding to $e_P \approx 0.65$ in the DU case) and (b) circuits with low entangling power ($J_z = 0.8$, corresponding to $e_P \approx 0.06$ in the DU case).
    The gray curve corresponds to the baseline: a Haar random matrix with dimensions $2^L$.
    We plot the IPC for the DU (blue circle) and non-DU (pink cross, $\Delta t = \pi/6$) ergodic circuits, as well as the DU (green circle) and non-DU (green cross) interacting integrable circuit [Eq.~\eqref{eq:XXZ_parametrization}].
    The numerics indicate that the DU circuits outperform their non-DU counterpart.
    In the bottom row, we plot the IPC as a function of noise for different circuit sizes ($L = 3, 5, 7, 9$ -- triangle, square, pentagon and star markers, respectively) and $J_z = 0.8$ (low DU entanglement).
    We show the results for the (c) non-DU and (d) DU ergodic circuits. 
    Panel (e) displays the Haar random benchmark. 
    Simulations for panels (a) and (b) were performed with $L = 7$, for $40 \times 50 = 2000$ task and circuit realizations, respectively.
    We train the model over $100L$ time steps in this figure.
    }
    \label{fig:capacity_vs_noise}
\end{figure}

\begin{figure*}
    \centering
    \includegraphics[width=.7\textwidth]{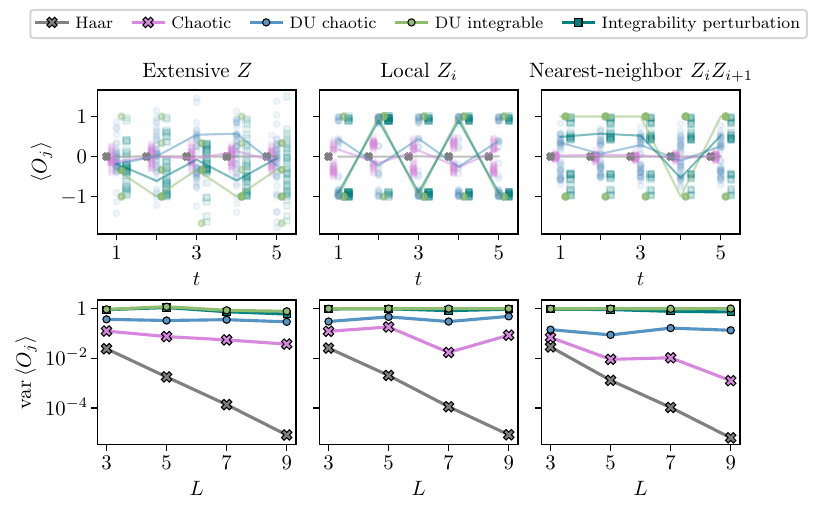}
    \caption{
    Exponential concentration in circuit models for different choices of observables.
    We plot the results for (a, d) total magnetization $\frac{1}{\sqrt{L}}\sum Z_i$, (b, e) a local Pauli operator in the middle of the chain $Z_i$ and (c, f) a nearest-neighbor operator $Z_i Z_{i + 1}$.
    We compute the expected values as a function of time for $40 \times 25 = 1000$ task and stochastic realizations, respectively.
    In the top row [panels (a) - (c)] we plot a subset of $5 \times 8 = 40$ of these realizations.
    We consider a Haar random reservoir (gray curve), the ergodic circuit (pink cross, $\Delta t = \pi/3$), the DU ergodic circuit (blue circle), the DU integrable circuit (green circle), and the circuit with a small perturbation from the integrable point (teal square). 
    We can observe that the Haar random reservoir fully concentrates all values around the mean (zero). 
    The ergodic circuit displays larger variance, which is further increased with dual-unitarity.
    The integrable circuit, on the other hand, corresponds to a discrete and well-spaced distribution.
    We show [bottom row, (d) - (f)] the variance as a function of system size, elucidating how different observables concentrate.
    The local structure of the brickwork circuits and, to a lesser extent, dual-unitarity and integrability, help avoid the hurdle of exponential concentration.  
    }
    \label{fig:concentration}
\end{figure*}

Our objective here is to simulate finite-shot noise in our simulations and to verify whether either integrability or dual-unitarity can assist in combating this problem.
The answer is affirmative in both cases, as we show in Fig.~\ref{fig:capacity_vs_noise}.
We model finite-shot noise by including Gaussian noise with standard deviation $\sigma_{\mathrm{Noise}} \sim 1/\sqrt{N}$, mimicking the fluctuations we would expect from a finite number of measurements $N$.
We repeat all the training steps discussed in Sec.~\ref{sec:QRC}, but this time we do so by sampling the random variable $\xi \sim \mathcal{N}(0, \sigma_{\mathrm{Noise}})$ from the Normal distribution. 
We then include it as a noise term in Eq.~\eqref{eq:features_graphical}, where the corresponding finite-shot feature reads $x_{t, j} \rightarrow x_{t, j} + \xi$.
This procedure is repeated for different values of $\sigma_{\mathrm{Noise}}$, which lead us to the results in Fig.~\ref{fig:capacity_vs_noise}, where we plot the IPC as a function of the finite-shot noise, for different circuit choices.

Basic results for exponential concentration are shown in Fig.~\ref{fig:capacity_vs_noise}~(a) and (b) for the NARMA($3$) task.
We consider two distinct scenarios: 
one where we take the entangling power to be large [panel (a)], choosing $J_z = 0.1$ [see Eq.~\eqref{eq:entangling_power}].
Conversely, we choose another configuration with low entangling power, choosing $J_z = 0.8$ [panel (b)].
As a basic benchmark, we take the reservoir to be a featureless model corresponding to a Haar random matrix of dimension $2^L$, agreeing with the circuit dimensionality.
While such a choice does not carry the same physical notion as the circuits we consider, it serves as a useful benchmark.
We clearly see that the Haar random reservoir not only performs worse in the absence of noise, but also exhibits a rapid degradation of the IPC with increasing $\sigma_{\mathrm{Noise}}$.
Once we consider the circuit models, we observe an improvement in all scenarios.
Numerically, the local architecture of the qubit circuits appears to mitigate exponential concentration.
Related concentration effects in temporal quantum learning models have recently been investigated in Ref.~\cite{xiongRoleScramblingNoise2025}.
Furthermore, when comparing the DU (circular markers) and non-DU (crosses) implementations, we see that the DU circuit displays higher IPC and is also more robust against noise.
This is more apparent in panel~(b), where the ergodic DU circuit performs much better than the non-DU one at large noise, $\sigma_{\mathrm{Noise}} \sim 0.1$.
Finally, integrability also appears to play an important role:
in almost all cases, the integrable circuit appears (i) to be more robust against finite-shot noise and (ii) to perform at least as well as the ergodic configuration.
The former is seen more clearly in panel~(a): even though the ergodic DU circuit (blue circle) initially performs better than the non-DU integrable one (green cross), as the noise increases the ergodic circuit quickly exhibits a drop in performance, while the integrable one remains robust for longer.
In Figs.~\ref{fig:capacity_vs_noise}~(c)-(e) we show complementary plots.
We once again display the IPC as a function of $\sigma_{\mathrm{Noise}}$, but this time for different circuit sizes $L$.
Surprisingly, in the non-DU case we do not observe large differences between system sizes.
In the DU case, meanwhile, the performance appears to be optimal for $L = 7$, with all curves converging at large noise.
Finally, in the Haar random case we see that the performance decreases markedly as the system size increases, and so does the tolerance to finite-shot noise.
Fig.~\ref{fig:capacity_vs_noise}~(e) makes the problem of exponential concentration particularly clear: the Haar reservoir has no locality or other mechanism that could protect it from exponential concentration, and the effect is strongest there.

In Fig.~\ref{fig:concentration} we show results that shed further light on this discussion and provide some intuition for these observations.
We consider a stochastic sequence of inputs, which we choose to be binary variables $s_t \in \{ 0, 1\}$, for simplicity (see Ref.~\cite{sanniaExponentialConcentrationSymmetries2025}). 
The idea in this figure is to simulate the dynamics of the reservoir under the erase-input protocol from Sec.~\ref{sec:QRC} for different circuit choices, and then verify how it responds to these inputs.
In the top row [Figs.~\ref{fig:concentration}~(a)-(c)], we plot the expected value of a chosen observable (shown on top of each column) as a function of time.
Here, we have considered that the reservoir already underwent the washout phase, so we consider an arbitrary sequence of five time steps. 
We consider configurations analogous to those shown in Fig.~\ref{fig:capacity_vs_noise}, but now include an extra curve for a circuit under a small integrability-breaking perturbation. 
This corresponds to choosing a small $\epsilon = 0.02$ in Eq.~\eqref{eq:gate_parametrization}.
The results are shown for $1000$ stochastic realizations, where stronger colors indicate a larger number of realizations with a given value (these plots can thus be seen as a ``top view" of a histogram).
What we immediately see is that the expected value concentrates very sharply around zero in the Haar random case.
In the ergodic case, we see that, although this distribution is centered around zero, it is much more spread out.
This is precisely what makes the brickwork architecture more resistant to exponential concentration: 
it takes longer for the finite-shot noise to become comparable to the typical expected values in the circuit.
Comparing the DU (blue circle) and non-DU (pink cross) cases makes this even clearer:
choosing the DU circuit increases the variance of the distribution and makes it more spread out, as seen in the corresponding densities.
Finally, this effect is clearest in the integrable cases: 
the expected values are concentrated around very distinct ``islands", reflecting the non-ergodic nature of the circuit.
In this case, it becomes much easier to distinguish the measured signals from noise.
The effect of the integrability perturbation is simply to spread out these individual ``islands", eventually merging them back into the ergodic case once the perturbation is strong enough.
These results are consistent with arguments of typicality in ergodic vs. integrable systems~\cite{DAlessio2016}.

As discussed in Ref.~\cite{thanasilpExponentialConcentrationQuantum2024}, one way to diagnose this effect is to compute the variance of these measurements.
The smaller the variance, the more concentrated they are.
As expected, in the Haar random case the variance decays exponentially with $L$.
However, in all other cases it does so at a much slower (non-exponential) rate over the system sizes and circuit depths considered here,
and the variance is clearly larger in the DU and integrable cases.
Although the scaling with $L$ is not fundamentally different from that of generic unitary circuits, the bottom row clearly shows that both dual-unitarity and integrability increase the variance and thereby shield the circuit against exponential concentration.


\section{Discussion and conclusion}

We have considered the use of dual-unitary quantum circuits as a platform for quantum reservoir computing.
Such circuits lead to both technical and conceptual advantages. 
For the former, dual-unitarity leads to an enhanced performance for short- to moderate delays, and results in an improved robustness to finite-shot noise. 
For the latter, the mechanisms of operator growth and solitons in dual-unitary circuits can be directly related to nonlinearity and memory effects in the reservoir. 
Dual-unitarity should also be supplemented with a moderate amount of entanglement generation, which we here quantify through the entangling power, in order to lead to an optimal performance.

The inclusion of a repeated-input scheme also allowed us to improve the control over nonlinear tasks.  
Our approach showed how the ergodic configuration of the circuits made them more agnostic to the fine-tuning of features, and thus better equipped to deal with such nonlinearities -- as we have clarified through our depiction of how temporal correlations are captured by operator growth. 
Moreover, our results also provided practical advantages in terms of shielding the reservoir against exponential concentration.
Although dual-unitarity in itself does not fundamentally avoid the concentration of measure in the readouts and its underlying \emph{scaling}, this construction led to a perceivable improvement in the distribution of expectation values of local observables at a constant prefactor.

We believe that our work can pave the way for future investigations in this direction.
Dual-unitarity is a rich and growing field, which has undergone substantial progress in the past decade, and the existing literature in its entirety can provide a much wider landscape than what we have presented so far.
For instance, general qudit-gates display a richer range of dynamical features, where entangling, ergodic and mixing properties can be more extensively tuned~\cite{ratherCreatingEnsemblesDual2020, aravindaDualunitaryQuantumBernoulli2021a}. 
However, this proves to be a challenge when compared to the current work because we also lose the simple and exhaustive parametrization of the DU gates used in Eq.~\eqref{eq:gate_parametrization}.
In such scenario, one would thus need to be more careful when exploring the relevant parameter space.
Nevertheless, there has been some recent interest in more applied aspects of dual-unitarity~\cite{riddellQuantumStateDesigns2025}, so such research directions can indeed prove to be promising. 
Indeed, studies in quantum information scrambling have gone into a very similar direction:
some recent investigations have very cleverly explored how information scrambling can lead to metrological advantage, in a very familiar language based on OTOCs and quantum circuits~\cite{kobrinUniversalProtocolQuantumEnhanced2024, geInformationScramblingEnhancedQuantumSensing2025, huQuantumEnhancedSensingEnabled2026a}.
Such type of adjacent work might be very helpful in understanding the future and potential of quantum reservoirs.
Similarly, generalizations of dual-unitarity, such as hierarchical DU  gates~\cite{yuHierarchicalGeneralizationDual2024, ramppInfiniteLevelHierarchySolvable2026}, can lead to more sophisticated circuits which can provide further insights into the interplay between information theoretic quantities~\cite{xiaQuantumMagicNoncommutativity2026a, keenanStorageScramblingLoss2026, dingThermodynamicsQuantumReservoir2026}, quantum dynamical behavior and learning properties in quantum reservoir computing.


\section*{Acknowledgments}

Both authors acknowledge support from the Max Planck Society. G.O.A. acknowledges helpful discussions with Edmilson Roque, Supanut Thanasilp and Jordi Riu Vicente. 


\onecolumngrid
\appendix

\section{Two-qubit gates in the Weyl chamber}

\subsection{Performance across the Weyl chamber}
\label{app:weyl}

\begin{figure*}
    \centering
    \includegraphics[width=0.9\textwidth]{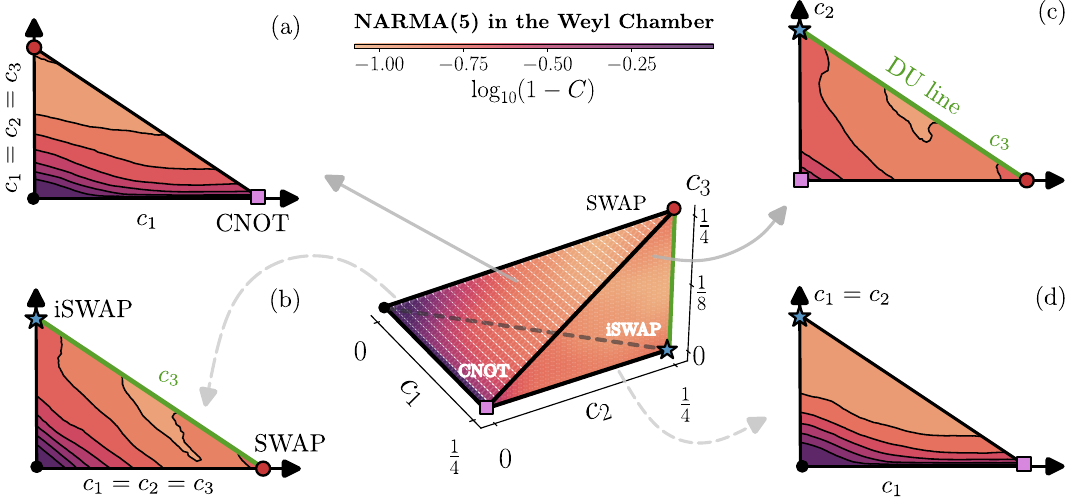}
    \caption{
    Plot of the error $1-C$ (in $\log_{10}$ scale) for the NARMA($5$) task across the Weyl chamber, following the Cartan parametrization in Eq.~\eqref{eq:cartan_full}.
    The gates of reference are the four vertices and their equivalence classes: purely local gates (black dot), CNOT gate (pink square), SWAP gate (red circle) and iSWAP gate (blue star).
    We highlight the dual-unitary line (green line), corresponding to the family of gates from Eq.~\eqref{eq:gate_parametrization}.
    We plot the results for (a) the locals-CNOT-SWAP face, (b) the XXZ family face, (c) the CNOT-SWAP-iSWAP face and (d) the XY family face.
    Each point in the density plots above corresponds to an average over $40 \times 25 = 1000$ task and gate realizations, respectively. 
    Simulations have been performed for $L = \tau = 5$.    
	Our results indicate that the optimal configuration for this task corresponds to a vicinity of gates around the DU line.
	Note that these correspond to the gates with moderate entangling power, per the discussion in Sec.~\ref{sec:benchmark_entanglement}.
    }
    \label{fig:weyl}
\end{figure*}

In the main text, we focused on understanding the role of dual-unitarity and integrability in the protocol.
This naturally made us consider a more restrictive choice of parameters.
In this section, we strive to better understand how the choice of two-qubit gates affects the reservoir in a more general manner, considering a more extensive parameter space instead. 
For that, we consider the Cartan decomposition as a generalized form of Eq.~\eqref{eq:XXZ_parametrization}, used to parametrize arbitrary two-qubit gates, i.e. SU(4) matrices ~\cite{khanejaCartanDecompositionSU2n2001a, zhangGeometricTheoryNonlocal2003, tucciIntroductionCartansKAK2005}.
We write such Cartan gates $U_{4 \times 4}$ as:
\begin{equation}\label{eq:cartan_full}
    U_{4 \times 4}
    =
    (u_+ \otimes u_-)
    V_{\mathrm{Cartan}}
    (v_+ \otimes v_-),
\end{equation}
where
\begin{equation}\label{eq:cartan}
    V_{\mathrm{Cartan}}
    =
    \exp\left[-i
    \left(
      c_1\, X \otimes X
    + c_2\,  Y \otimes Y
    + c_3\, Z \otimes Z
    \right)\right],
\end{equation}
parametrizes the non-local part of the gate, and $u_+, u_-, v_+, v_- \in SU(2)$ correspond to single-qubit gates. 
Due to symmetry considerations, it suffices to consider gates with $\pi/4 \geq c_1 \geq c_2 \geq c_3 \geq 0$~\cite{zhangGeometricTheoryNonlocal2003}.
By choosing $c_1 = c_2 = \pi/4$ and $c_3 =  \pi J_z/4$ the non-local part $V_{\mathrm{Cartan}}$ recovers the DU XXZ gate of Eq.~\eqref{eq:XXZ_parametrization}. 
The parameters $c_i$, with $i = 1, 2, 3$, are called the \emph{non-local content} of the gate.
This means that all gates with the same non-local content, following Eq.~\eqref{eq:cartan_full}, are equivalent under local unitary transformations, i.e. they are said to belong to the same equivalence class. 
Therefore, the question we want to probe is: 
how does the choice of parameters $\boldsymbol{c} = (c_1, c_2, c_3)$ affect the IPC?
This decomposition enjoys a convenient geometric depiction in terms of the so-called \emph{Weyl chamber}~\cite{balakrishnanCharacterizingGeometricalEdges2009, mandarinoBipartiteUnitaryGates2018a, ratherConstructionLocalEquivalence2022, linLetEachQuantum2023, hahnAbsenceLocalizationWeakly2024a}.
If we further consider the invariance of entangling power upon complex conjugation, this restriction confines the parameters to a three-dimensional tetrahedron, shown in Fig.~\ref{fig:weyl}.
We can associate each point in the tetrahedron, corresponding to $\boldsymbol{c}$, with an equivalence class of non-local gates (see, e.g., the discussion in Sec.~B of Ref.~\cite{mullerOptimizingEntanglingQuantum2011a}).
The vertices of this tetrahedron parametrize the non-local part of the Cartan decomposition of familiar gates:
\begin{enumerate}
    \item At (0, 0, 0) we have that $V_{\mathrm{Cartan}}$ reduces to the identity. This choice thus corresponds to the set of products of single-qubit gates (black dot in Fig.~\ref{fig:weyl}).
    \item At ($\pi/4$, 0, 0) we recover the CNOT gate (pink square).
    \item At ($\pi/4$, $\pi/4$, 0) we obtain the iSWAP gate (blue star).
    \item Finally, the vertex ($\pi/4$, $\pi/4$, $\pi/4$) corresponds to the SWAP gate (red circle).
\end{enumerate}
We plot our results for the NARMA($5$) task in Fig.~\ref{fig:weyl} across the four faces of the Weyl chamber.
For each data point, we average both over random realizations of the NARMA task, as well as over random choices of SU($2$) gates in Eq.~\eqref{eq:cartan_full}, i.e. we also average over different local realizations of the gates, while keeping the nonlocal part  [Eq.~\eqref{eq:cartan}] fixed.
We then plot the error $1 - C$, on a logarithmic scale, at the corresponding point in the Weyl chamber. 
For a complete visualization, we plot the 2D projection of each of the faces in panels (a) - (d).
An accompanying figure [Fig.~\ref{fig:weyl_chamber_ep}] can be found in App.~\ref{app:weyl_chamber_ep}, where we plot the entangling power [Eq.~\eqref{eq:entangling_power_general}] for a general two-qubit gate (along the surface of the Weyl chamber).
We now proceed to discuss the different faces in Fig.~\ref{fig:weyl}.

\paragraph{Locals-CNOT-SWAP face.}

This face is shown in Fig.~\ref{fig:weyl}~(a) and contains the line parametrizing the XXX family of gates with $c_1 = c_2 = c_3$.
We observe that the reservoir performs badly around the identity, as one would expect: generalization in such cases is very limited, and the reservoir is typically unable to capture some nonlinear effects or to retain memory for time-lags larger than system size.
A similar bad performance is seen for the CNOT gate, in spite of the fact that it displays maximal entangling power of $2/3$~\cite{mannaEntanglingPowerGate2024}.
This indicates that maximizing entanglement itself is not enough to make the reservoir sufficiently expressive.
Moving closer to the dual-unitary SWAP gate generally results in an improved performance, which drops down exactly at the SWAP gate (which cannot generate entanglement).
    
\paragraph{XXZ family face.}

This corresponds to the face where $c_1 = c_2$, and contains the DU line, which we have parametrized in the previous sections according to Eq.~\eqref{eq:gate_parametrization}.
This highlighted green line corresponds to the configurations studied in Fig.~\ref{fig:tasks_vs_entanglement}.
We recover the expected result that the performance is optimal for the segment of the DU line away from the vertices, as well as for  more generic non-DU gates in the nearby vicinity of this line.

\paragraph{CNOT-SWAP-iSWAP face.}

Shown in Fig.~\ref{fig:weyl}~(c). This face is parametrized by choosing $c_1 = \pi/4$.
We observe a very similar behavior here, with a local minimum along the DU line around intermediate values of $c_3$, fully analogous to the XXZ family face. 

\paragraph{XY family.}

Shown in Fig.~\ref{fig:weyl}~(d), it corresponds to the bottom face of the tetrahedron (as drawn in the figure) and is determined by setting $c_3 = 0$.
Here, we find something qualitatively similar to the Locals-CNOT-SWAP face [Fig.~\ref{fig:weyl}~(a)].
Overall, our analysis across the Weyl chamber indicates that the best-performing gates concentrate in the vicinity of the DU line. Notably, maximally entangling gates such as CNOT and iSWAP underperform, consistent with the results of Sec.~\ref{sec:benchmark_entanglement}, where gates with low-to-moderate entangling power were found to perform best.

\subsection{Entangling power across the Weyl chamber}
\label{app:weyl_chamber_ep}

\begin{figure*}[t!]
    \centering
    \includegraphics[width=0.9\textwidth]{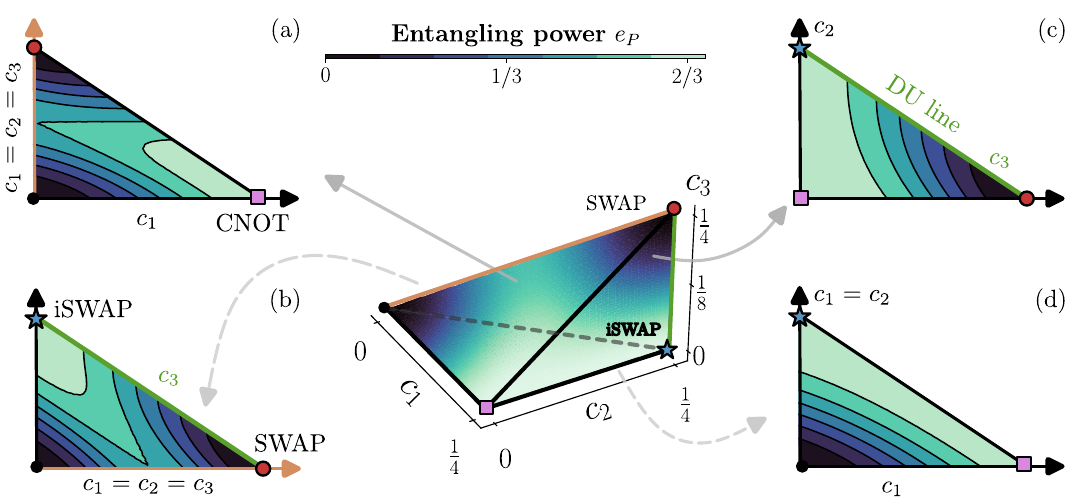}
    \caption{
    Plot of the entangling power as a function of the parameters $c_1$, $c_2$, and $c_3$ across the Weyl chamber.
    This figure is analogous to Fig.~\ref{fig:weyl}, and all details are the same.
    For completeness, we further highlight the line $c_1 = c_2 = c_3$ with an orange hue, denoting the family of fractional SWAP gates.
    The entangling power has been normalized such that it lies between $0$ and $2/3$.
    Both the CNOT (pink square) and iSWAP (blue star, locally equivalent to DCNOT) display maximum entangling power.
    Conversely, note that the entangling power vanishes for local (black circle) and SWAP gates (red circle).
    }
    \label{fig:weyl_chamber_ep}
\end{figure*}

\begin{figure*}
    \centering
    \includegraphics[width=0.45\columnwidth]{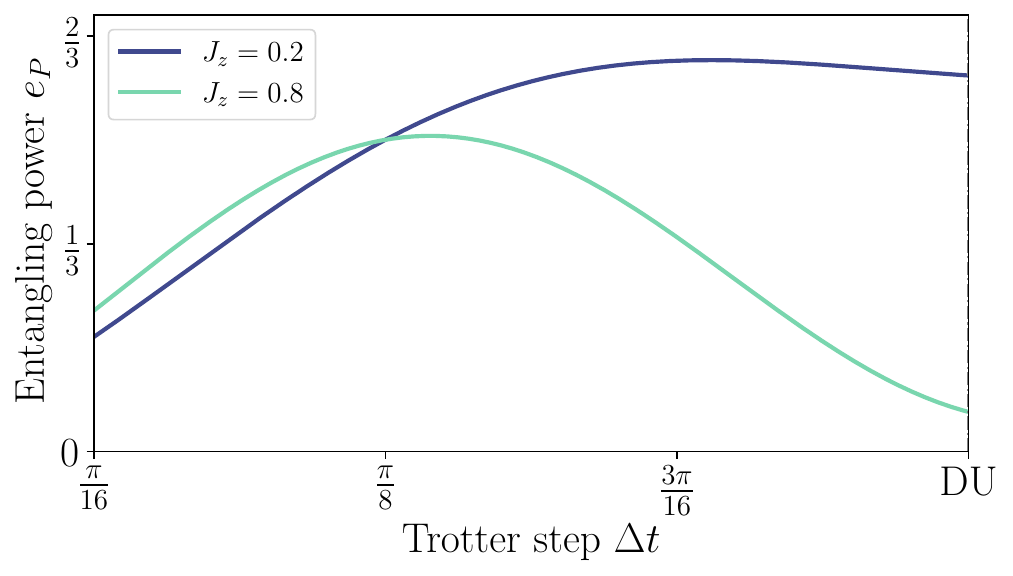}
    \caption{
    Entangling power $e_P$ as a function of the Trotter step $\Delta t$ according to Eq.~\eqref{eq:entangling_power_dt}.
    The curves are shown for $J_z = 0.2$ and $J_z = 0.8$, which we call the high- and low-entangling regimes.
    Compare Figs.~\ref{fig:tasks_vs_trotter} and Figs.~\ref{fig:tasks_vs_trotter_small_Jz} to see how different choices of $J_z$ and $\Delta t$ can strongly affect the performance, especially in regard to the optimality (or lack thereof) at the DU point.
    }
    \label{fig:entangling_power_vs_trotter}
\end{figure*}

For a general two-qubit gate, parametrized by the Cartan coordinates $\boldsymbol{c} = (c_1, c_2, c_3)$ as in Eq.~\eqref{eq:cartan}, the entangling power (with the linear entropy as the entanglement measure) takes the closed form~\cite{balakrishnanEntanglingPowerLocal2010}:
\begin{equation}\label{eq:entangling_power_general}
    e_P(c_1,c_2,c_3)
    =
    \frac{1}{6}\Big[\,3-\big(\cos 4c_1\cos 4c_2+\cos 4c_2\cos 4c_3+\cos 4c_3\cos 4c_1\big)\Big],
\end{equation}
normalized such that its maximum value is $2/3$.
Specializing to the Trotterized XXZ gate of Eq.~\eqref{eq:XXZ_parametrization}, which corresponds to $c_1 = c_2 = \Delta t$ and $c_3 = J_z\, \Delta t$, this reduces to an explicit two-parameter expression in the Trotter step $\Delta t$ and the anisotropy $J_z$:
\begin{equation}\label{eq:entangling_power_dt}
    e_P(\Delta t, J_z)
    =
    \frac{1}{6}\Big[\,3-\cos^2(4\Delta t)-2\cos(4\Delta t)\cos(4 J_z \Delta t)\Big].
\end{equation}
This expression is associated with the simplified decomposition from Eq.~\eqref{eq:cartan}, given in terms of Trotter step and anisotropy only.
Numerical results for this parametrization are given in Fig.~\ref{fig:entangling_power_vs_trotter}.
Note that up to $\Delta t = \pi/8$ the two curves corresponding to $J_z = 0.2$ and $J_z = 0.8$ follow each other very closely.
Afterwards, the gate with smaller anisotropy ($J_z = 0.2$) overtakes and becomes much more entangling.
This gap becomes largest at the DU point. 
This might explain why we see the largest differences there.
At the dual-unitary point $\Delta t = \pi/4$, this recovers the usual expression Eq.~\eqref{eq:entangling_power}, i.e. $e_P = \frac{2}{3}\cos^2\!\left(\frac{\pi}{2} J_z\right)$ for DU qubit gates.
In Fig.~\ref{fig:weyl_chamber_ep} we plot Eq.~\eqref{eq:entangling_power_general} in the Weyl chamber and juxtapose it with Fig.~\ref{fig:weyl}.
Note how we find maximum entanglement at the CNOT and iSWAP vertices, even though these gates do not perform particularly well for most tasks, and as we have seen, this is especially so around the CNOT vertex.
Similarly, the entangling power vanishes around the family of locals and SWAP gates, as these are obviously non-entangling.
Finally, around the $\sqrt{\mathrm{SWAP}}$ gate, midway through the SWAP and the non-locals (with $c_1 = c_2 = c_3 = \pi/8$), we have an intermediate amount of entanglement -- similar to the optimal DU gates we have found.

\section{Further investigations on the role of the Trotter step}
\label{app:trotter_high_entanglement}

\begin{figure}
    \centering
    \includegraphics[scale=0.5]{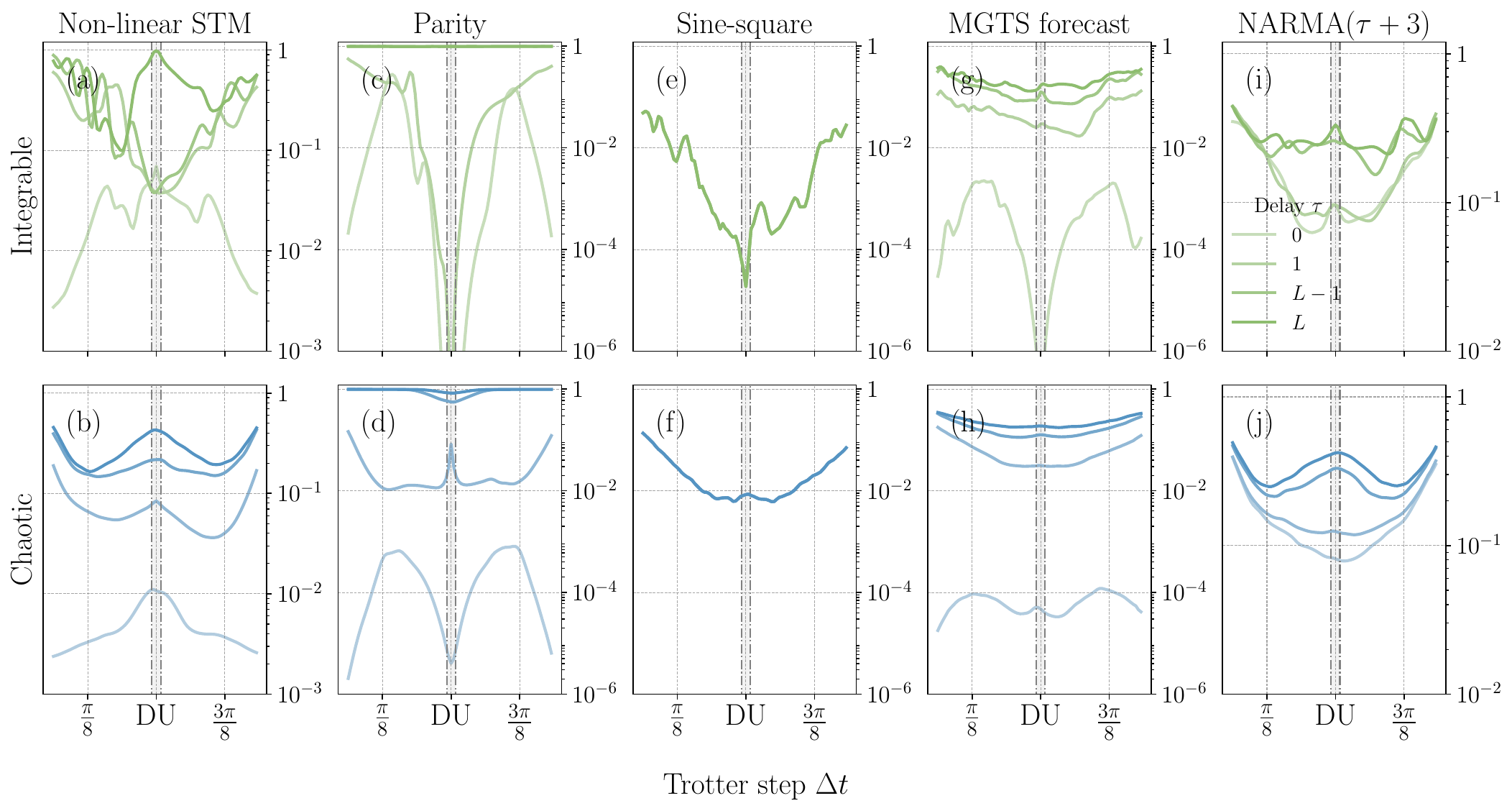}
    \caption{
    We perform numerical investigations analogous to Fig.~\ref{fig:tasks_vs_trotter}.
    We take $J_z = 0.1$ (with entangling power $e_P \approx 0.65$), corresponding to a highly entangling gate.
    All other parameters are kept the same.
    In this regime, the relation between dual-unitarity and optimality is not as straightforward as in Fig.~\ref{fig:tasks_vs_trotter}.
    In this case, this will depend more strongly on the task and delay considered.
    }
    \label{fig:tasks_vs_trotter_small_Jz}
\end{figure}

One might wonder how different the results from Fig.~\ref{fig:tasks_vs_trotter} would be in the regime of small anisotropy $J_z$, which translates into large entangling power according to Eq.~\eqref{eq:entangling_power}.
Indeed, we should be careful when considering such a scenario, as the results in Fig.~\ref{fig:tasks_vs_trotter_small_Jz} indicate.
Exactly as in the discussion in Sec.~\ref{sec:optimality_DU}, we separate the plots into the top row (green curves) and bottom row (blue curves), corresponding to integrable and ergodic dynamics.
Similarly, in each of the columns we display a different task.
We first turn our attention to the first column, with the results plotted for the nonlinear STM task.
The first thing worth highlighting is the fact that, contrary to the regime of low entanglement in Fig.~\ref{fig:tasks_vs_trotter}, the DU point in this task might actually lead to a local \emph{maximum} for the error ($1 - C$).
This happens for a few delay values in the integrable case, and for all of them in the chaotic case.
We actually do see similar trends in the chaotic parity check task (d) and for the chaotic MGTS task (h), although less extremely.
In the sine-square task, meanwhile, and for the integrable MGTS task, the DU point retains its optimality.
The NARMA task [panels (i) and (j)] is similar to the nonlinear STM task, and the optimality of the DU neighborhood depends on the delay considered.
Overall, this figure shows that although dual-unitarity may lead to moderate improvements in a few cases, whether that is really the best choice is still very task-dependent. 
Similar considerations hold when it comes to entanglement, as we see in Fig.~\ref{fig:tasks_vs_entanglement}.

\section{Choice of features}
\label{app:choice_features}

\begin{figure}[t]
    \centering
    \includegraphics[scale=.5]{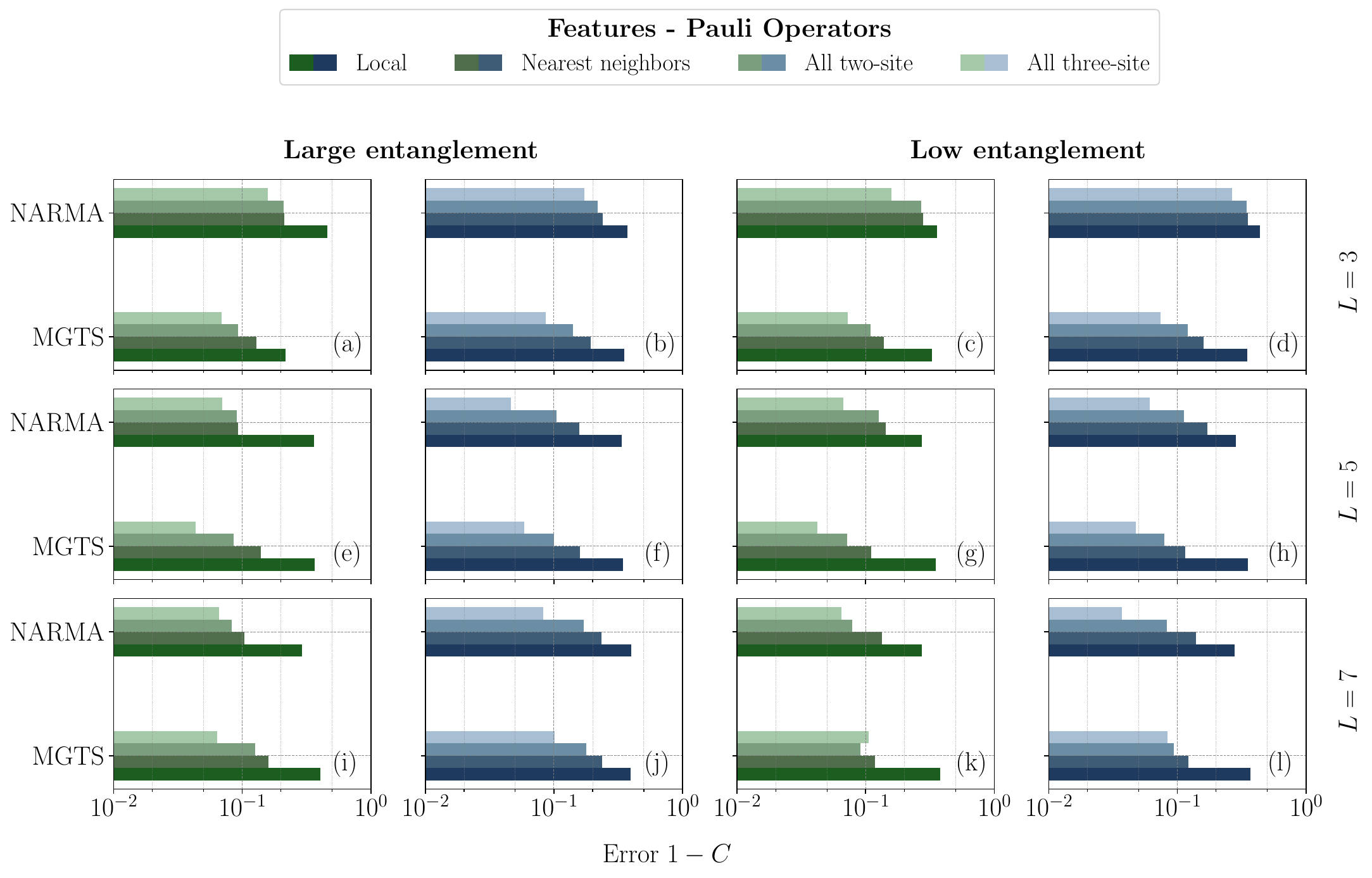}
    \caption{
    Bar plot showing the error $1 - C$ for different tasks (MGTS and NARMA), for both integrable (green) and chaotic (blue) circuits and $\tau = L + 1$.
    Different bars correspond to different choices of features, with lighter colors corresponding to an increasingly larger number of features.
    We show results for both high- and low-entanglement scenarios, in analogy with Fig.~\ref{fig:capacity_vs_noise}, corresponding to $J_z = 0.1$ and $J_z = 0.8$, respectively.
    Each row, from top to bottom, corresponds to $L = 3, 5, 7$, respectively (right-hand side labels).
    Simulations were performed using $50 \times 40 = 2000$ gate and task realizations, respectively.
    Increasing the number of features generally improves the performance, leading to smaller errors.
    Due to the large number of features we consider here, these simulations were run for $280L$ and $60L$ training and validation steps, respectively. 
    This avoids an undetermined linear system during the regression step. 
    }
    \label{fig:features_app}
\end{figure}

Another aspect that can be explored is how the choice of features affects the performance of the reservoir.
In other words, one can ask the following question:
to what extent does the inclusion of either a larger number of nodes or more non-local operators improve the predictions?
We show results in Fig.~\ref{fig:features_app}.
We consider both integrable (green) and chaotic (blue) circuits, shown in alternating columns. 
We also consider high- and low-entanglement reservoirs, each corresponding to a pair of columns.
Finally, each of these rows corresponds to a different system size, labeled on the right-hand side, and different bars denote different choices of Pauli operators as features.
These features are chosen as:
(i) the scenario with purely local single-site Pauli operators,
(ii) the one with the inclusion of nearest-neighbor operators $X_{i}X_{i + 1}$, $Y_{i}Y_{i + 1}$ and $Z_{i}Z_{i + 1}$,
(iii) the one where we further include \emph{all} two-site Pauli operators over all sites and Pauli labels, and finally
(iv) the scenario where we also include all three-site Pauli operators over all sites and Pauli labels.
The choices (i) - (iv) correspond to the different shades of green and blue in the figure, with the darkest and lightest tones corresponding to (i) and (iv), respectively.

Let us consider panel~(a) of Fig.~\ref{fig:features_app}.
Note the log-scale in the horizontal axis.
We can see, for both tasks, that a larger number of features also leads to improved performance, as one would naturally expect.
This effect is, of course, more noticeable for larger system sizes (compare panels (a) and (i), for instance), as the number of available features becomes much larger as we increase $L$.
Similarly, this is also very dependent on the task and reservoir type. 
Let us take the MGTS task for system size $L = 7$ and a low entanglement reservoir, as depicted in panels (k) and (l).
There, we can see that the improvement from local to nearest neighbor features is larger in the ergodic case than in the integrable case.
Conversely, the jump from nearest-neighbor features to all two-site operators is much more significant in the integrable case, as shown in panel (k).
Nevertheless, it is remarkable that the reservoir displays satisfactory performance even when we operate with a small number of nodes and very local features.

Moreover, when comparing the columns in the last two rows, namely (c, g, k) and (d, h, l), respectively, which correspond to the \emph{low}-entanglement regime, we can see that increasing the system size $L$ monotonically improves the performance.
We should be careful to note that this is not always the case, however:
for the collection of parameters considered here, the monotonic improvement with $L$ breaks down in the highly entangled chaotic reservoir.
When comparing panels (f) and (j) for the MGTS task, for example, we see that the reservoir with $L = 5$ actually performs better.
These results indicate that the optimal reservoir might depend both on other details of the reservoir and on \emph{which} features are available.
The conclusion is that one should therefore consider this type of trade-off:
if we cannot access a sufficiently large number of features or fine-tune the remaining parameters of the reservoir, increasing its size is not necessarily the best strategy.

\twocolumngrid
\bibliography{library}

\end{document}